\def\notes{0}
\documentclass{article}
\usepackage{hyperref}
\usepackage{graphicx} 
\usepackage{amssymb,amsmath,amsthm}
\usepackage{thmtools,thm-restate,mathtools}
\usepackage{multirow} 
\usepackage{xcolor,xspace}
\definecolor{mygreen}{RGB}{34,164,31}
\usepackage{bbm}
\usepackage{fullpage}
\usepackage{url}
\usepackage{cleveref}
\usepackage{enumitem}
\setlist{topsep=1mm, itemsep=0mm, leftmargin=6mm} 
\usepackage{bbm}
\hypersetup{
    colorlinks=true,
    linkcolor=blue,
    anchorcolor = red, 
    citecolor = mygreen,
    filecolor=magenta,      
    urlcolor=cyan}

    \title{Fully Dynamic Graph Algorithms with Edge Differential Privacy}

\author{Sofya Raskhodnikova\footnote{Boston University, Boston, MA, USA. Email: \texttt{sofya@bu.edu}. S.R.\ was supported in part by the NSF award DMS-2022446.} \and Teresa Anna Steiner\footnote{IMADA, University of Southern Denmark, Odense, Denmark. Email: \texttt{steiner@imada.sdu.dk}. T.A.S. was supported by a research grant (VIL51463) from VILLUM FONDEN.}}
\date{}

\newcommand{\Alg}{\mathcal{A}}
\newcommand{\range}{\mathrm{range}}
\newcommand{\graphseq}{\mathcal{S}}
\newcommand{\diffedge}[1]{dE_{#1}}
\newcommand{\diffedgeprime}[1]{dE_{#1}'}
\newcommand{\diffedgeplus}[1]{dE^+_{#1}}
\newcommand{\diffedgeminus}[1]{dE^-_{#1}}
\newcommand{\diffedgeprimeplus}[1]{d{E^+_{#1}}'}
\newcommand{\diffedgeprimeminus}[1]{d{E^-_{#1}}'}
\newcommand{\allgraphs}[1]{\mathcal{G}(#1)}
\newcommand{\symdif}[2]{\Delta(#1,#2)}
\newcommand{\out}{\mathrm{Out}}

\newcommand{\Lap}{\mathrm{Lap}}

\newtheorem{theorem}{Theorem}[section]

\newtheorem{definition}[theorem]{Definition}
\newtheorem{lemma}[theorem]{Lemma}
\newtheorem{corollary}[theorem]{Corollary}
\newtheorem{remark}[theorem]{Remark}

\newtheorem{claim}[theorem]{Claim}

\newtheorem{example}[theorem]{Example}
\counterwithin{figure}{section}

\newcommand{\R}{\mathbb{R}} 
\newcommand{\N}{\mathbb N}

\newcommand{\eps}{\varepsilon}

\def\polylog{\operatorname{polylog}}

\newcommand{\cC}{\mathcal{C}}

\newcommand{\cY}{\mathcal{Y}}
\newcommand{\uni}{\mathcal{U}}

\newcommand{\ededgedp}{$(\eps,\delta)$-edge-DP\xspace}
\newcommand{\eedgedp}{$\eps$-edge-DP\xspace}
\newcommand{\reddp}[1]{$D$-rest. $(\eps,\delta)$-edge-DP\xspace}
\newcommand{\redp}[1]{$D$-rest. $\eps$-edge-DP\xspace}
\newcommand{\degr}[1]{\mathrm{deg}(#1)}

\newcommand{\innerprod}[2]{\mathrm{InnerProd}(#1,#2)}
\newcommand{\marginals}[2]{\mathrm{Marginals}(#1,#2)}
\newcommand{\degreelist}{{\sf DegreeList}\xspace}
\newcommand{\trianglecount}{{\sf TriangleCount}\xspace}
\newcommand{\highdegree}[1]{\mathrm{HighDegree}(#1)}
\newcommand{\degreelistmath}{f_{\mathrm{deg}}}
\newcommand{\highdegreemath}[1]{f_{\geq #1}}
\newcommand{\trianglecountmath}{f_{\triangle}}
\newcommand{\degreehist}{{\sf DegreeHist}\xspace}
\newcommand{\degreehistmath}{f_{\mathrm{degHist}}}

\newcommand{\mincut}{{\sf MinCut}\xspace}
\newcommand{\mincutmath}{f_{MC}}
\newcommand{\maxmatch}{{\sf MaximumMatching}\xspace}
\newcommand{\maxmatchmath}{f_{MM}}
\newcommand{\conncomp}{{\sf ConnectedComponents}\xspace}
\newcommand{\conncompmath}{f_{CC}}
\newcommand{\edgecount}{{\sf EdgeCount}\xspace}
\newcommand{\edgecountmath}{f_{\mathrm{edges}}}

\ifnum\notes=1
    \newcommand{\sr}[1]{{\color{blue} #1}}
    \newcommand{\ts}[1]{{\color{mygreen} #1}}
    \newcommand{\srnote}[1]{{\color{blue}\footnote{{\color{blue} {\bf SR:} #1}}}}
    \newcommand{\tsnote}[1]{{\color{mygreen}\footnote{{\color{mygreen} {\bf TS:} #1}}}}

\else 
    \newcommand{\srnote}[1]{}
    \newcommand{\tsnote}[1]{}
    
    \newcommand{\sr}[1]{{#1}}
    \newcommand{\ts}[1]{{#1}}

\fi

\begin{document}

\maketitle
\begin{abstract}

    We
study differentially private algorithms for analyzing 
    graphs 
in the challenging setting of continual release with fully dynamic updates, where edges are inserted and deleted over time, and the algorithm is required to update the solution at every time step. Previous work has presented differentially private algorithms for many graph problems that can handle insertions only or deletions only (called {\em partially dynamic algorithms}) and obtained some hardness results for the fully dynamic setting. The only algorithms in the latter setting were for the edge count, given by Fichtenberger, Henzinger, and Ost (ESA '21), and for releasing the values of all graph cuts, given by Fichtenberger, Henzinger, and Upadhyay (ICML '23). We provide the first differentially private and fully dynamic graph algorithms for several other fundamental graph statistics (including the triangle count, the number of connected components, the size of the maximum matching, and the degree histogram), analyze their error, and show strong lower bounds on the error for all algorithms in this setting. Previously, only lower bounds for 
\sr{\emph{purely} differentially private algorithms}
were known; our lower bounds give an exponential improvement in terms of the dependence on the number of time steps, while applying to algorithms 
\sr{satisfying \emph{pure} as well as \emph{approximate} differential privacy.}

We study two variants of {\em edge differential privacy} for fully dynamic graph algorithms: {\em event-level} and {\em item-level}. Under the former notion, two graph
update sequences are considered neighboring  if, roughly speaking, they differ in at most one update; under the latter notion, they can differ only in updates pertaining to one edge. Differential privacy requires that for any two neighboring inputs, the output distributions of the algorithm are close. We give upper and lower bounds on the error of both---event-level and item-level---fully dynamic algorithms for several fundamental graph problems. No fully dynamic algorithms that are private at the item-level (the more stringent of the two notions) were known before. In the case of item-level privacy, for several problems, our algorithms match our lower bounds.
\end{abstract}
\thispagestyle{empty}

\newpage
\clearpage
\pagenumbering{arabic} 


\section{Introduction}\label{sec:intro}

Graph
algorithms
provide important tools for understanding networks and relationships.
Specifically, graphs can be used to model complex relational databases, and analyzing those has applications in recommender systems, analysis of social networks, market analysis, etc. \cite{pods/Leskovec23}.
However, some 
    graphs
contain sensitive personal information, 
that must be protected when results of statistical analyses performed on them are published.
Differential privacy~\cite{DworkMNS16} has emerged as the standard for rigorous privacy guarantees. It has been adapted to graph data and widely investigated in the {\em static setting}, also referred to as the {\em batch setting}, where the input  
    graph 
does not change (see \cite{RaskhodnikovaS16-E} for a survey of differentially private analysis of graphs in this setting). An additional challenge in studying 
    graphs 
that capture information about people is that they evolve over time. Differential privacy was first studied in the setting of {\em continual release} (also called {\em continual observation}), where data changes over time and published statistics must be continually updated, by Dwork et al.~\cite{DworkNPR10} and Chan et al.~\cite{ChanSS11}. This difficult setting was adapted to graph data and investigated by 
Song et al.~\cite{SongLMVC18}, Fichtenberger et al.~\cite{FichtenbergerHO21}, and Jain et al.~\cite{JainSW}. In this setting, the algorithm receives a sequence of 
updates, one per time step, where each update is either an insertion or a deletion of an edge. {\em Fully dynamic} 
algorithms handle both insertions and deletions, whereas {\em partially dynamic} 
algorithms handle insertions only or deletions only.
In this work, we study fully dynamic graph algorithms in the continual release model. 

Differential privacy (DP), intuitively, guarantees that, for any two neighboring 
datasets,
the output distributions of the algorithm are roughly the same. We consider {\em edge DP,} introduced in ~\cite{NissimRS07}, that uses the notion of {\em edge-neighboring graphs}. Two graphs are edge-neighboring if they differ in one edge. (There is also a stronger notion of {\em node DP}, first studied in~\cite{BlockiBDS13,KasiviswanathanNRS13,ChenZ13}.) Differential privacy is defined with two parameters, $\eps$ and $\delta$; see Definitions~\ref{def:dp} and~\ref{def:DP-dynamic-graphs}. When $\delta=0$, it is referred to as {\em pure DP}; the setting when $\delta>0$ is called {\em approximate DP}.

The continual release model 
comes with two natural definitions of neighboring sequences and two corresponding variants of differential privacy: {\em event-level} and {\em item-level}, and we study both of them. Update sequences are {\em event-neighboring} if they differ in one update; they are {\em item-neighboring} if they differ on an arbitrary number of updates pertaining to one item. 
For edge DP, we adapt these definitions as follows.
Two graph sequences are considered {\em event-level edge-neighboring}
if one can be obtained from the other by replacing either one update with a no-op or two consecutive updates on the same edge with no-ops.\footnote{\label{footnote:explanation-of-two-changes}We consider two updates on the same edge $e$  consecutive if the updates between them  do not involve $e$. Consecutive updates on the same edge must be of different types: an insertion and a deletion (in either order). We allow up to {\em two} updates to be replaced, as opposed to one in the classical continual release setting, for a technical reason: since an edge can be inserted only when it is absent and deleted only when it is present, there exist update sequences for which changing any one update results in an invalid graph sequence. This happens, for example, if the updates alternate between inserting and deleting the same edge, ending in an insertion. As a result, our definition is more natural in the dynamic setting.} 
%
Two graph sequences are {\em item-level edge-neighboring} if they are the same on all updates, except updates on one edge.
Event-level privacy is less stringent than item-level privacy.
Note that there is no distinction between event-level and item-level edge DP for partially dynamic algorithms, because each edge can be updated at most once. But the picture is more nuanced for fully dynamic algorithms.

Our goal is to investigate the best error achievable by fully dynamic, 
edge differentially private (edge-DP) graph algorithms. We call an algorithm $(\alpha,\beta)$-accurate if, with probability at least $1-\beta$, it has additive error at most $\alpha$ at every time step. For algorithms that output real-valued vectors, the additive error is measured in terms of $L_\infty$, i.e., as the maximum over all coordinates of the vector. We express our error bounds in terms of the number of nodes in the graph, denoted by $N$; the number of time steps, called the {\em time horizon} and denoted by $T$; and sometimes also in terms of the bound $D$ on the maximum degree. 

The systematic
investigation by Fichtenberger et al.~\cite{FichtenbergerHO21} provided partially dynamic private algorithms with additive error $\polylog T$ for numerous graph problems. The partially dynamic setting under node DP was further investigated by Jain et al.~\cite{JainSW}. The only fully dynamic graph algorithms in previous work were for the edge count, given by Fichtenberger et al.~\cite{FichtenbergerHO21}, and for releasing the values of all graph cuts, given by Fichtenberger et al.~\cite{FichtenbergerHU23}. Both algorithms satisfy event-level, edge DP; the former algorithm has dependence $O(\log^{3/2}T)$ on the time horizon $T$ in the additive error, whereas the latter algorithm has dependence $O(N^{3/2}\sqrt{\log N}\log^{3/2}T)$ on $T$ and the number of nodes $N$.
Fichtenberger et al.~\cite{FichtenbergerHO21} also obtained $\Omega(\log T)$ lower bounds for many graph problems in this setting and lower bounds in terms of $N$ (that hold for large $T$) in the item-level DP setting. Their lower bounds are only for {\em pure DP}, that is, when $\delta=0$. Fichtenberger et al.~\cite{FichtenbergerHU23} 
show a lower bound on the accuracy of continually releasing the value of the min cut with event-level, edge DP, but in a different model: they have parallel edges and allow multiple edges to be changed at every time step, both of which make the problem harder and lead to a lower bound on the error that is larger than our event-level upper bound for the value of min cut.

\subsection{Our results}\label{sec:results}
Our results on  event-level and item-level edge-DP are summarized in Tables~\ref{table:results-event-level} and \ref{table:results-item-level}, respectively.

 We give the {\em first  fully dynamic,} differentially private algorithms for several fundamental graph statistics, including triangle count ($\trianglecountmath$), the number of connected components ($\conncompmath$), the size of the maximum matching ($\maxmatchmath$), 
 the number of nodes of degree at least $\tau$ ($\highdegreemath{\tau}$), the degree histogram ($\degreehistmath$), and the degree list ($\degreelistmath$) with edge DP at both event level and item level. For the edge count ($\edgecountmath$), we give the first item-level edge-DP algorithm. (See \Cref{def:specific-graph-functions} for problem definitions.) Our event-level algorithm for $\degreelistmath$
 has additive error $\alpha$ polylogarithmic in $T$ and $N$. All other algorithms have error polynomial in these parameters.


\setlength\tabcolsep{6pt}
\begin{table}[t]
    \begin{tabular}{|l||c|c|c|c|}
    \hline
    & \multirow{2}*{Upper bounds} & \multirow{2}*{Lower bounds} & Lower bounds & \cite{FichtenbergerHO21}\\
    &&&out-det& $\delta=0$\\
    \hline
    \hline
   \multirow{2}*{$\trianglecountmath$} & $\tilde{O}(\min(T,\sqrt{TN},\sqrt[3] TN,N^3
   ))$ & ${\Omega}(\min(T,\sqrt[3] TN^{2/3},N^2))$& same as  & \multirow{4}*{$\Omega(\log T)$}
     \\
      
     & \Cref{thm:upper_triangle}  &\Cref{thm:lower_triangle_submatrix} &lower bounds &\\
    \cline{1-4}
    
     $\conncompmath$
     & $\tilde{O}(\min(\sqrt[3] T,N))$ &$ \Omega(\min(\sqrt[4] T,\sqrt{N})$ & $\tilde{\Omega}(\min(\sqrt[3] T,N))$ & \\ 
   $\ldots$ & \Cref{table:results-item-level}&\Cref{cor:lb-event-level}& \Cref{cor:lb-event-level-outputd}&\\
     \hline
      \multirow{2}*{$\degreelistmath$} & $O(\log T\log (TN))$ &  $\Omega(\log T)$ &  same as& \multirow{2}*{---}\\ 
     &  \Cref{lem:estimating_deg}&\Cref{remark:deg-list}&lower bounds&\\
     \hline
    \end{tabular}
    \caption{Bounds on error $\alpha$ for $(\alpha,\beta)$-accurate, {\bf event-level} \ededgedp fully dynamic graph algorithms
    (suppressing factors polynomial in $\frac 1\eps\log \frac{1}{\delta\beta}$). We use $\tilde{O}(X)$ and $\tilde{\Omega}(X)$ to hide $\polylog X$ factors. The bounds under $\conncompmath...$ hold for $\conncompmath, \maxmatchmath,\highdegreemath{\tau},\degreehistmath$.
    }\label{table:results-event-level}
\end{table}

\setlength\tabcolsep{0.8pt}
\begin{table}[t]
    \begin{tabular}{|c||c|c||c|c|c|}
    \hline
    & \multicolumn{2}{|c|}{Approximate DP ($\delta>0)$} & \multicolumn{3}{|c|}{Pure DP ($\delta=0)$} \\ \cline{2-6} 
    & Upper bounds & Lower bounds &  Upper bounds & Lower bounds & \small{\cite{FichtenbergerHO21}}\\
    \hline
    \hline
    \multirow{2}*{$\trianglecountmath$} & \small{$\tilde{O}(\min(T^{\frac 43},\sqrt[3] TN, N^3
    ))$} & \small $\tilde{\Omega}(\min(T^{\frac 7 6},\sqrt[3] TN,N^3)
    )$& \small $\tilde{O}(\min(T^{\frac 3 2},\sqrt T N, N^3
    ))$ & \small $\tilde{\Omega}(\min(T^{\frac 5 4},\sqrt T N,N^3)
    )$ &  $\Omega(N^3)$\\
    &\Cref{cor:upperbound_triangle_item-level} & \Cref{thm:lb-triangle_item-level}&\Cref{cor:upperbound_triangle_item-level}&\Cref{thm:lb-triangle_item-level}&\\
    \hline
    
     $\conncompmath$
     &   & $\tilde{\Omega}(\min(\sqrt[3] T,N))$ &&$\tilde \Omega(\min(\sqrt T,N))$ & $\Omega(N)$\\ 
     $\cdots$&$\tilde{O}(\min(\sqrt[3] T,N))$ & \Cref{cor:lb-item-level}&$\tilde O(\min(\sqrt T,N))$ &\Cref{cor:lb-item-level}&
     \\\cline{1-1}\cline{3-3}\cline{5-6}
     \multirow{2}*{$\degreelistmath$} &\Cref{cor:ub-low-sensitivity-item-level}&as for $\conncompmath$&\Cref{cor:ub-low-sensitivity-item-level}&as for $\conncompmath$&\multirow{2}*{---}\\
     &  & \Cref{remark:deg-list} && \Cref{remark:deg-list} &
     \\
     \hline
     \multirow{2}*{$\edgecountmath$} & $\tilde O(\min(\sqrt[3] T,N^2))$ & $\Omega(\min(\sqrt[3] T,N^2))$& $\tilde O(\min(\sqrt T,N^2))$&$\Omega(\min(\sqrt T,N^2))$&$\Omega(N^2)$\\
     &\Cref{cor:ub-low-sensitivity-item-level}  & \Cref{thm:lb-edge_item-level} &\Cref{cor:ub-low-sensitivity-item-level}  & \Cref{thm:lb-edge_item-level}&\\
     \hline
    \end{tabular}
    \caption{Bounds on error $\alpha$ for $(\alpha,\beta)$-accurate,  {\bf item-level} \ededgedp fully dynamic graph algorithms 
    (suppressing factors polynomial in $\frac 1\eps\log \frac{1}{\delta\beta}$. We use $\tilde{O}(X)$ and $\tilde{\Omega}(X)$ to hide factors polylogarithmical in $X$).  The bounds under $\conncompmath..$ hold for $\conncompmath, \maxmatchmath,\highdegreemath{\tau},\degreehistmath$. Bounds from \cite{FichtenbergerHO21} hold only for sufficiently large $T$. 
    }\label{table:results-item-level}
    
\end{table}

\paragraph{Event-level.} We 
demonstrate that 
$\trianglecountmath,\conncompmath,\maxmatchmath,\highdegreemath{\tau}$, and $\degreehistmath$ are fundamentally different from \edgecount and \degreelist by showing that every event-level edge-DP algorithm for these five problems must have additive error polynomial in $T$ and $N$ (as opposed to polylogarithmic in these parameters), thus providing an exponential improvement on the previously known lower bounds of $\Omega(\log T)$ by \cite{FichtenbergerHO21} in terms of the dependence on the time horizon $T$. Our lower bounds apply for general $\delta\in[0,1)$, that is, even for approximate DP, whereas the lower bounds in \cite{FichtenbergerHO21} hold only for pure DP. Specifically, we show that when $N$ is sufficiently large, the best additive error of event-level edge-DP algorithms must be $\Theta(T)$ for $\trianglecountmath$ and between  $O(T^{1/3})$ and $\Omega(T^{1/4})$ for $\conncompmath,\maxmatchmath,\highdegreemath{\tau},\degreehistmath$.

\paragraph{Event-level output-determined.} While there remains an intriguing gap between upper and lower bounds for these problems, we are able to eliminate it for $\conncompmath,\maxmatchmath,\highdegreemath{\tau},\degreehistmath$ for a special class of algorithms, called {\em output-determined (out-det)}, first considered in \cite{HenzingerSS24}, where the authors showed a lower bound for output-determined algorithms for counting the number of distinct elements in a stream. Output-determined algorithms have roughly the same output distributions on inputs with the same 
output sequences (see \Cref{def:output-determined}.) 
All known algorithms for these four problems (all of which come from our work) are output-determined. Our lower bounds for output-determined algorithms demonstrate that we would need a different approach to improve the error.

\paragraph{Item-level.} We also give item-level lower bounds on the additive error for fully dynamic graph algorithms. For all problems listed in the table, except for the triangle count, our item-level lower bounds match the error of our algorithms up to polylogarithmic factors. Our bounds for triangle count under item-level $(\eps,\delta)$-edge-DP are tight (up to polylogarithmic factors) for $N\leq T^{5/6}$, and under item-level $\eps$-edge-DP for $N\leq T^{3/4}$. Further, 
we show that when $N$ is sufficiently large, the additive error of item-level $(\eps,\delta)$-edge-DP algorithms must be between $O(T^{4/3})$ and $\Omega(T^{7/6})$ for $\trianglecountmath$ and   $\tilde\Theta(T^{1/3})$ for  $\conncompmath,\maxmatchmath,\highdegreemath{\tau},\degreehistmath, \edgecountmath$, and $\degreelistmath$ when $\delta>0$. When $\delta=0$, it is between $T^{3/2}$ and $T^{5/4}$ for $\trianglecountmath$ and $\Theta(\sqrt{T})$ for the remaining problems.


\paragraph{Discussion and Bounded Degree.} Putting our results together with the $O(\log^{3/2} T)$ event-level upper bound for $\edgecountmath$ by \cite{FichtenbergerHO21}, we see that $\edgecountmath$ and $\degreelistmath$ become much more difficult in the item-level setting: the error grows from $\polylog T$ to $T^{1/2}$ or $T^{1/3}$ (depending whether we consider pure or approximate DP).
For the remaining problems, while $\conncompmath,\maxmatchmath,\highdegreemath{\tau},\degreehistmath$ behave similarly in terms of the additive error of fully dynamic algorithms, the triangle count stands out. We further investigate it in the event-level setting by considering error bounds in terms of the maximum degree $D$ and time horizon $T$. We prove the upper bound of 
$\tilde{O}(\min(\sqrt{TD},\sqrt[3] TD)$ and the lower bound of ${\Omega}(\min(\sqrt[3] TD^{2/3},\sqrt[4] TD))$ for this problem.
We stress that our algorithms are differentially private even when the degree bound is violated, but the accuracy guarantees expressed in terms of $D$ rely on the input graph being of degree at most $D$ at every time step. Subroutines that are differentially private only when the input graphs have maximum degree at most $D$ are called {\em $D$-restricted}. As shown by Jain et al.~\cite{JainSW}, such a subroutine can exhibit blatant failures of differential privacy on graphs that do not satisfy the promise if they are used as stand-alone algorithms.

\subsection{Our Techniques}\label{sec:techniques}
We prove the first superlogarithmic lower bounds for dynamic differentially private algorithms for graph problems.
Our starting point for proving lower bounds is the sequential embedding technique proposed by Jain et al.~\cite{JainRSS23} and further developed by Jain et al.~\cite{JainKRSS23}. The main idea is to reduce from a problem in the batch setting that returns multiple outputs about the same individuals and use different time steps to extract answers that correspond to different outputs. In particular, \cite{JainRSS23} used a reduction from the 1-way marginals problem in the batch setting and then applied lower bounds of~\cite{HardtT10} and~\cite{BunUV18} for releasing all 1-way marginals; \cite{JainKRSS23} used a reduction from the Inner Product problem in the batch setting and applied the lower bounds for this problem from~\cite{DinurN03,DworkMT07,MirMNW11,De12}.
 
\paragraph{Lower bounds via Submatrix Queries.} To prove our event-level lower bounds for $\trianglecount$ (stated in \Cref{thm:lower_triangle_submatrix}), we reduce from the Submatrix Query problem, which has not been used in the context of continual release before. It was defined by Eden et al.~\cite{EdenLRS23} for studying the local model of differential privacy, where each party holds their own private data and interacts with the rest of the world using differentially private algorithms. In the submatrix query problem, the input 
dataset
is an $n\times n$ matrix $Y$ of random bits, each belonging to a different individual. Each submatrix query is specified by two vectors $a,b\in\{0,1\}^n$. The answer to the query $(a,b)$ is $a^TYb=\sum_{i=1}^{n}\sum_{j=1}^n a[i]Y[i,j]b[j]$. A submatrix query is a type of a {\em linear query}. The answer to a linear query is a dot product of the secret 
dataset
(viewed as a vector; in our case, of length $n^2$) and a query vector of the same length (in our case, a vectorization of the outer product of vectors $a$ and $b$). Dinur and Nissim \cite{DinurN03}, in the paper that laid foundations for establishing the field of differential privacy, show that answering many random linear queries too accurately leads to blatant privacy violations. Eden et al.~\cite{EdenLRS23} extended their attack to submatrix queries. Implicitly, they showed differentially private algorithms for answering $\Omega(n^2)$ submatrix queries on $n\times n$ 
dataset must have additive error $\Omega(n)$. We use this lower bound together with our reduction from Submatrix Queries to triangle counting with event-level edge-DP.

Our reduction transforms a
dataset
 $Y$ and a set $Q$ of submatrix queries  into a dynamic graph sequence~$\graphseq$. It ensures that the answers to different submatrix queries in $Q$ correspond to triangle counts in $\graphseq$ at different time steps. Moreover, the transformations of neighboring 
datasets
$Y$ and $Y'$ (with the same query set $Q$) return event-level, edge-neighboring graph sequences. Our reduction encodes the secret 
dataset
$Y$ as edges of a bipartite graph. Then it crucially uses insertions and deletions by first inserting and then deleting edges that encode each query. The main advantage of using submatrix queries instead of arbitrary linear queries in our reduction is that they can be encoded more efficiently, i.e., with fewer edges. This creates shorter dynamic graph sequences, leading to stronger lower bounds in terms of the time horizon $T$.

\paragraph{General lower bound framework.} We provide a general lower bound framework for fully dynamic graph algorithms, which we employ to prove nearly all of our lower bounds (with the notable exception of event-level triangle counting, for which we use the submatrix queries method described above, in order to get a better bound). The framework is based on the existence of a small graph gadget with two special edges such that the value of the target function (e.g., $\maxmatchmath$) remains the same when only one of the special edges is removed, but changes by a specific amount when both special edges are removed. We call such a gadget {\em 2-edge distinguishing for $f$} (see \Cref{fig:gadgets_gen}). We show how to use a 2-edge distinguishing gadget for $f$ to obtain a reduction from
the Inner Product problem (see \Cref{def:inner-product}) in the batch setting to  the continual release of $f$ in the fully dynamic, event-level setting. We also show how to use a slightly simpler gadget for $f$ to obtain a reduction from
1-way Marginals (see \Cref{def:marginals}) in the batch setting  to the continual release of $f$ in the fully dynamic, item-level setting. Finally, 2-edge distinguishing gadgets are used in the general reduction from 1-way Marginals to get event-level lower bounds in the output-determined setting. We then apply our general framework to obtain numerous lower bounds for specific problems.

\paragraph{Transformation from $D$-restricted to general algorithms in the event-level setting.} One of the most common methods for obtaining DP graph algorithms is to start by designing an algorithm that is tailored to a specific graph family, typically graphs of degree at most $D$. The reason for this is that some graph statistics have much lower sensitivity on bounded degree graphs. E.g., if one edge in an $N$-node graph is added or removed, the triangle count can change by $N-2$; however, in a graph of degree at most $D$, it can change by at most $D-1$. In the batch setting, many techniques (notably,  based on graph projections \cite{BlockiBDS13,KasiviswanathanNRS13,DayLL16}  and Lipschitz extensions \cite{BlockiBDS13,KasiviswanathanNRS13,ChenZ13,RaskhodnikovaS16,RaskhodnikovaS16-E,DayLL16,KalemajRST23}) have been developed for transforming algorithms tailored to bounded-degree graphs to algorithms that are private on {\em all graphs} without noticeably increasing their error on bounded-degree graphs.
For dynamic graph algorithms,  $D$-restricted algorithms were studied in \cite{SongLMVC18,FichtenbergerHO21}. For the insertions-only setting, Jain et al.~\cite{JainSW} present a general projection that 
transforms a $D$-restricted  algorithm into an algorithm which is differentially private on the universe of \emph{all} input graph sequences, with similar error guarantees depending on the maximum degree of the input sequence (up to log factors). 

We obtain a similar result in the fully dynamic setting for event-level edge DP: i.e., if for all $D$ there exists an event-level $D$-restricted edge-DP algorithm for estimating a function on all dynamic graph sequences of maximum degree $D$, with an error bound depending on $D$, then there exists an event-level edge-DP algorithm on \emph{all} dynamic graph sequences, with an error bound depending on the maximum degree of the sequence. The error bounds are the same as for the $D$-restricted case, up to factors logarithmic in $TN$ and small dependence on the privacy parameters and the failure probability $\beta$.

One of the tools in our transformation is our event-level algorithm for degree list, which is based on the continual histograms algorithm from \cite{FichtenbergerHU23}.  We use the degree-list algorithm to keep a running estimate of the maximum degree. We run the corresponding degree restricted mechanism until the maximum degree increases significantly. We then re-initialize a new restricted mechanism with a higher degree bound.

\paragraph{Fully dynamic algorithm for triangle count.} Our event-level algorithm for triangle count uses our transformation from $D$-restricted to general algorithms. To get a $D$-restricted algorithm, we use a binary-tree-based mechanism for counting on the difference sequence for $\trianglecountmath$ (i.e., a sequence that records the change in $\trianglecountmath$ between time steps) and then add noise using the Gaussian mechanism. This strategy is used for other graph problems in \cite{FichtenbergerHO21} with Laplacian noise instead of Gaussian noise.
To bound the error of our mechanism, we give a careful analysis of the $L_2$-sensitivity of the counts stored in the binary tree.

\subsection{Additional Related 
Work}\label{sec:prior-work} The two concurrent works~\cite{DworkNPR10, ChanSS11} that initiated the investigation of the continual release model, also proposed the binary-tree mechanism for computing sums of bits, a.k.a. continual counting. The problem of continual counting was further studied in~\cite{DworkNRR15,FichtenbergerHU23,HenzingerUU23,sp/DongLY23,CohenLNSS}.
The binary-tree mechanism has been extended to work for sums of real values \cite{PerrierAK19}, weighted sums \cite{BolotFMNT13}, counting distinct elements \cite{BolotFMNT13, EpastoMMMVZ23, Ghazi0NM23,JainKRSS23,HenzingerSS24} and, most relevantly, partially dynamic algorithms for graph statistics \cite{FichtenbergerHO21}.

The first lower bound in the continual release model was an $\Omega(\log T)$ bound on the additive error of continual counting, shown by \cite{DworkNPR10}. The first lower bounds that depended polynomially on $T$ were proved by Jain et al.~\cite{JainRSS23} and applied to the problems of releasing the value and index of the attribute with the highest sum, given a 
dataset
that stores whether each individual possesses a given attribute. They pioneered the sequential embedding technique that reduces multiple instances of a static problem to a dynamic problem. Like in that paper, we also reduce from the 1-way marginals problem to obtain some of our lower bounds. However, our lower bound for the triangle count is proved by a reduction from a different problem, and our reductions use the specific structure of the graph problems we consider.

First edge-DP algorithms appeared in~\cite{NissimRS07}; they handled the cost of the minimum spanning tree and triangle count. Since then differentially private graph algorithms have been designed for a variety of tasks, including estimating subgraph counts \cite{BlockiBDS13, KasiviswanathanNRS13, ChenZ13, KarwaRSY14, DingZB018, LiuML20}, degree list \cite{KarwaS12}, degree and triangle distributions \cite{HayLMJ09,RaskhodnikovaS16,DayLL16,LiuML20}, the number of connected components \cite{KalemajRST23}, spectral properties \cite{WangWW13,AroraU19}, parameters in stochastic block models \cite{BorgsCS15, BorgsCSZ18, SealfonU21}; training of graph neural networks \cite{DaigavaneMSTAJ21}; generating synthetic graphs \cite{KarwaS12,ZhangNF20}, and many more~\cite{BlockiBDS12,GuptaRU12,Upadhyay13,WangWZX13,ProserpioGM14,LuM14, ZhangCPSX15, MulleCB15, NguyenIR16, RoohiRT19,ZhangN19,AhmedLJ20,BlockiGM22,DLRSSY22}.

\paragraph{Concurrent work.} Recently and independently, \cite{EpastoLMZ} showed similar lower bounds for fully dynamic algorithms achieving event-level, $\eps$-edge-DP via a reduction from InnerProduct; however,  the focus of their paper is not on the fully dynamic model, but on obtaining better additive error in the insertions-only model. In the fully dynamic model, they prove $\Omega\Big(\min\Big(\sqrt{\frac{N}{\eps}},\frac{T^{1/4}}{\eps^{3/4}},N,T\Big)\Big)$ lower bounds for $\maxmatchmath$, $\conncompmath,$ and $\trianglecountmath$. Their lower bounds match ours from Corollary~\ref{cor:lb-event-level} for $\maxmatchmath$ and $\conncompmath$ in terms of the dependence on $N$ and $T$, but also have an explicit dependence on $\eps$ in them for the special case of pure differential privacy (ours are stated for constant $\eps$ and more general $(\eps,\delta)$-DP).
However, our lower bound for $\trianglecountmath$ from Theorem~\ref{thm:lower_triangle_submatrix} is better by a polynomial factor in $T$ and $N$.

\subsection{Discussion and Open Questions}
Our transformation from $D$-restricted to general edge-DP algorithms works in the event-level setting. An interesting open question is if one can design such a transformation for the item-level setting (with a small error blowup).

Our error bounds for $\conncompmath,\maxmatchmath,\highdegreemath{\tau}$, and $\degreehistmath$ exhibit similar dependence on $T$ as the bounds in \cite{JainKRSS23} on counting the number of distinct elements  in  turnstile streams in the continual release model. Unlike in our setting, where an edge can be added only if it is present and deleted only if it is absent, turnstile streams might have an element added or deleted no matter what the current count for this element is. Only elements with positive counts are included in the current distinct element count. Interestingly, \cite{HenzingerSS24} show that when elements can only be added when absent and deleted when present (i.e., in the so-called ``likes'' model), the additive error drops down to polylogarithmic in $T$. We conjecture that improving the bounds for counting distinct elements in the event-level setting would lead to better bounds for graph problems as well.

\subsection{Organization of Technical Sections}

We start by giving definitions and the background on differential privacy and continual release in \Cref{sec:prelims}. In \Cref{sec:tringle-counting}, we collect all our results on triangle counting that require specialized techniques. \Cref{sec:algorithmic-tools} presents general algorithmic tools: the transformation from degree-restricted to general event-level algorithms in \Cref{sec:transformation-from-degree-restricted-to-private}, and algorithms based on recomputation at regular intervals in \Cref{sec:recomputing-strategy}. It also applies these tools to get most upper bounds on the error presented in Tables~\ref{table:results-event-level} and \ref{table:results-item-level}. Finally, \Cref{sec:lb-framework} presents our lower bound framework and most lower bounds on the error from the tables.

\section{Preliminaries}\label{sec:prelims}
We use $\log$ to denote the logarithm base 2 and $\ln$ to denote the natural logarithm. Let $[T]$ denote the set $\{1,2,\dots, T\}$. Let $\uni$ be a universe of data items and $n\in \N$. A 
\emph{dataset over $\uni$ 
of size $n$} 
is an $n$-tuple of elements from~$\uni$. We use $\uni^*$ to denote the set of all 
(of all sizes) over $\uni$.

\begin{definition}[Neighboring datasets]
Two datasets $x,y\in\uni^n$  are \emph{neighboring}, denoted $x\sim y$, if there is an $i\in[n]$ such that $x[i]\neq y[i]$, and $x[j]=y[j]$ for all $j\in[n]\setminus\{i\}$.
\end{definition}

\begin{definition}[Differential privacy \cite{DworkMNS06,DworkKMMN06}]\label{def:dp}
Let $\Alg$ be an algorithm which takes as input a 
dataset
over $\uni$. Let $\eps>0$ and $\delta\in[0,1)$. Algorithm $\Alg$ is {\em $(\eps,\delta)$-differentially private ($(\eps,\delta)$-DP)} if for all neighboring 
datasets
$x,y\in\uni^*$
and all $\out\subseteq \range(\Alg)$, 
\begin{align*}
    \Pr[\Alg(x)\in \out]\leq e^{\eps}\Pr[\Alg(y)\in \out]+\delta.
\end{align*}
If $\Alg$ is $(\eps,\delta)$-differentially private for $\delta=0$, it is also called $\eps$-differentially private ($\eps$-DP).
\end{definition}

\begin{definition}[Sensitivity] Let $k\in\mathbb{N}$, and $f:\uni^{*}\rightarrow \mathbb{R}^k$ 
be a function. Let $p\in\{1,2\}$.
The \emph{$L_p$-sensitivity of $f$}, denoted by $\Delta_p$, is defined as
    $\Delta_p=\max_{x\sim y}\|f(x)-f(y)\|_p.$
\end{definition}

\begin{definition}[Laplace distribution]\label{def:laplace}
The \emph{Laplace distribution} centered at $0$ with scale $b$ is the distribution with probability density function 
$
    f_{\Lap(b)}(x)=\frac{1}{2b}\exp\left(\frac{-|x|}{b}\right)$.
 We use $Y\sim \Lap(b)$ or just $\Lap(b)$ to denote a random variable $Y$ distributed according to $f_{\Lap(b)}(x)$.
\end{definition}

\begin{lemma}[Laplace Mechanism~\cite{DworkMNS06}] \label{lem:Laplacemech} Let $k\in\N$ and $\eps>0$ and 
$f:\uni^{*}\rightarrow \mathbb{R}^k$ 
be a function
with $L_1$-sensitivity $\Delta_1$. 
The Laplace mechanism is defined as
$\Alg(x)=f(x)+(Y_1,\dots,Y_k)$, where $Y_i \sim 
\Lap(\Delta_1/\eps)$ are independent random variables for all $i\in[k]$. The Laplace mechanism is $\eps$-differentially private. Further, for every $x\in\uni^{*}$ and every $\beta\in(0,1)$, it satisfies $\|\Alg(x)-f(x)\|_{\infty}\leq \frac{\Delta_1}\eps\ln\frac k \beta$ with probability at least $1-\beta$.
\end{lemma}
\begin{definition}[Normal Distribution] The \emph{normal distribution} centered at $0$ with variance $\sigma^2$ is the distribution with the probability density function
\begin{align*}
f_{N(0,\sigma^2)}(x)=\frac{1}{\sigma\sqrt{2\pi}}\exp\left(-\frac{x^2}{2\sigma^2}\right)
\end{align*}
We use $Y\sim N(0,\sigma^2)$ or sometimes just $N(0,\sigma^2)$ to denote a random variable $Y$ distributed according to $f_{N(0,\sigma^2)}$.
\end{definition}
 
\begin{lemma}[Gaussian mechanism~\cite{BlumDMN05,BunS16}]\label{lem:gaussianmech}
 Let $k\in\mathbb{N}$ and $f:\uni^{*}\rightarrow \mathbb{R}^k$ 
be a function  with $L_2$-sensitivity $\Delta_2$.
Let $\eps\in(0,1)$, $\delta\in(0,1)$, $c^2>2\ln(1.25/\delta)$, and $\sigma\geq c\Delta_2/\eps$. 
The Gaussian mechanism is defined as $\mathcal{A}(x)=f(x)+(Y_1,\dots,Y_k)$, where $Y_i\sim N(0,\sigma^2)$ are independent random variables for all $i\in[k]$. The Gaussian mechanism
is
$(\eps,\delta)$-differentially private. Further, for every $x\in \uni^{*}$, every $\beta\in(0,1)$, and $\sigma=\sqrt{2\ln(2/\delta)}\Delta_2/\eps$, it satisfies $\|\Alg(x)-f(x)\|_{\infty}\leq \frac{2\Delta_2}\eps \sqrt{\ln(2/\delta)\ln(2k/\beta)}$ with probability at least $1-\beta$.
\end{lemma}

\begin{lemma}[Gaussian tail bound]\label{lem:gaussian_tail}
Let $Y\sim N(\mu,\sigma^2)$ and $t\geq 0$. Then
    $\Pr[|Y-\mu|\geq \sigma t]\leq 2e^{-t^2/2}$.

\end{lemma}

\begin{lemma}[Simple composition~\cite{DworkL09,DworkRV10,DworkKMMN06}]\label{lem:composition_theorem} Let $\eps_1,\eps_2> 0$ and $\delta_1,\delta_2\in[0,1)$. 
Let $\Alg_1$ be an $(\eps_1,\delta_1)$-differentially private algorithm $\uni^{*}
\rightarrow \mathrm{range}(\Alg_1)$ and $\Alg_2$ an $(\eps_2,\delta_2)$-differentially private algorithm $\uni^{*}
\times\mathrm{range}(\Alg_1)\rightarrow \mathrm{range}(\Alg_2)$. Then $\Alg_1\circ \Alg_2$ is $(\eps_1+\eps_2,\delta_1+\delta_2)$-differentially private. 
\end{lemma}

\subsection{Graph Sequences}
For a set of nodes $V$, let $\allgraphs{V}$ denote the set of all simple graphs on $V$. The symmetric difference of two sets $S$ and $S'$, denoted $\symdif{S}{S'}$, is $(S\setminus S')\cup (S'\setminus S)$.
\begin{definition}[Edge-neighboring] 
Graphs $G=(V,E)$ and $H=(V,E')$ are \emph{edge-neighboring}, denoted by $G\sim H$, if $E$ and $E'$ differ in at most one element, i.e., $|\symdif{E}{E'}|\leq 1$.
\end{definition}

\begin{definition}[Dynamic graph sequence]\label{def:dynamic-graph-sequence}
    Let $V$ be a set of nodes and $T\in \mathbb{N}$. For all $t\in\{0\}\cup[T]$, let $G_t$ be a graph $(V,E_t)$ with some edge set $E_t\subseteq V\times V$ and $E_0=\emptyset$. Then $\graphseq=(G_1,\dots, G_T)$ is a \emph{dynamic graph sequence of length $T$} if 
    $|\symdif{E_{t-1}}{E_t}|\leq 1$ for all $t\in[T]$.
That is, every two consecutive graphs in the sequence differ by at most one edge insertion or deletion. 

To denote the edge difference sets, for all $t\in[T]$, we use $\diffedgeminus{t}=E_{t-1}\setminus E_{t}$ and $\diffedgeplus{t}=E_t\setminus E_{t-1},$ as well as
$\diffedge{t}=\diffedgeminus{t}\cup \diffedgeplus{t}=\symdif{E_t}{E_{t-1}}$.

We say $\graphseq$ has {\em maximum degree 
$D$} if $G_t$ has maximum degree 
$D$ for all $t\in[T]$.
\end{definition}

\paragraph{Continual release graph algorithms} Let $V$ be a set of nodes and $T,k\in\mathbb{N}$. A \emph{continual release graph algorithm} $\Alg:\allgraphs{V}^T\rightarrow (\mathbb{R}^k)^T$ receives as input\footnote{We assume $T$ is given to the algorithm. By a standard reduction \cite{ChanSS11}, all our error bounds hold if $T$ is not known in advance, up to polylogarithmic factors in $T$.} a dynamic graph sequence $\graphseq=(G_1,\dots,G_T)$ and, at every time step $t\in[T]$, produces an output $a_t=\Alg(G_1,\dots, G_t)\in \mathbb{R}^k$.

\begin{definition}[Item-level and event-level neighbors] Let $V$ be a set of nodes and $T\in \mathbb{N}$. Let $\graphseq$ and $\graphseq'$ be two dynamic graph sequences of length $T$. Let $\diffedgeplus{t}$, $\diffedgeminus{t}$ and $\diffedge{t}$ be the edge difference sets as in \Cref{def:dynamic-graph-sequence}, and $\diffedgeprimeplus{t}$, $\diffedgeprimeminus{t}$ and $\diffedgeprime{t}$ be the corresponding sets for $\graphseq'$. We say $\graphseq$ and $\graphseq'$ are \emph{item-level, edge-neighboring} if they differ in updates pertaining to at most one edge. That is, there exists an edge $e^*\in V^2$ such that $\diffedgeplus{t}\setminus\{e^*\}=\diffedgeprimeplus{t}\setminus\{e^*\}$ and $\diffedgeminus{t}\setminus\{e^*\}=\diffedgeprimeminus{t}\setminus\{e^*\}$ for all $t\in[T]$.

 We say $\graphseq$ and $\graphseq'$ are \emph{event-level, edge-neighboring} if they differ in at most one edge insertion (or deletion) and the next deletion (or insertion) of the same edge.\textsuperscript{\footnotesize{\em\ref{footnote:explanation-of-two-changes}}} 
 That is, there exists an edge $e^*\in V^2$ and a time interval $[t_1,t_2]$ with $t_1\in[1,T]$ and $t_2\in[2,T+1]$ 
 such that\footnote{To represent the case when $e^*$ is not updated after time step $t_1$, we set $t_2$ to $T+1$. In that case, the requirements for this time step in \Cref{item:EL-requirements} should be ignored.}  
\begin{enumerate}
    \item $\diffedgeplus{t}=\diffedgeprimeplus{t}$ and $\diffedgeminus{t}=\diffedgeprimeminus{t}$ for all $t\in[T]\setminus\{t_1,t_2\}$.
    \item $e^*\notin \diffedge{t}$ for all $t\in(t_1,t_2)$.
    \item\label{item:EL-requirements} 
    a)  $\diffedgeplus{t_1}=\diffedgeminus{t_2}=\{e^*\}$ and $\diffedgeprime{t_1}=\diffedgeprime{t_2}=\emptyset$ or 
    b) $\diffedgeminus{t_1}=\diffedgeplus{t_2}=\{e^*\}$ and $\diffedgeprime{t_1}=\diffedgeprime{t_2}=\emptyset$, 
\end{enumerate}
or if the symmetric properties are true with the roles of $\graphseq$ and $\graphseq'$ switched.
\end{definition}
Note that 
 if $\graphseq=(G_1,\dots, G_T)$ and $\graphseq'=(G'_1,\dots, G'_T)$ \sr{are event-level edge-neighboring, then they are also item-level edge-neighboring; if they are item-level edge-neighboring, then} the graphs $G_t$ and $G'_t$ are edge-neighboring for every $t\in [T]$.

\begin{definition}[Item-level and event-level edge-DP]\label{def:DP-dynamic-graphs} 
Let $V$ be a set of nodes and $T,k\in\mathbb{N}$. Let $\Alg$ be a continual release graph algorithm $\Alg:\allgraphs{V}^T\rightarrow(\mathbb{R}^k)^T$. Let $\eps>0$ and $\delta\geq 0$. The algorithm $\Alg$ is item-level (respectively, event-level), $(\eps,\delta)$-edge-DP, if for all item-level (respectively, event-level) edge-neighboring dynamic graph sequences $\graphseq\in\allgraphs{V}^T$ and $\graphseq'\in\allgraphs{V}^T$  and all $\out\subseteq(\mathbb{R}^k)^T:$
\begin{align}\label{eq:dp}
    \Pr\left[(\Alg(G_1,\dots,G_t)_{t\leq T}\in \out\right]\leq e^{\eps}\Pr\left[(\Alg(G'_1,\dots,G'_t)_{t\leq T}\in \out\right]+\delta.
\end{align}
We use $\eps$-edge-DP as a shorthand for $(\eps,0)$-edge-DP.
\end{definition}
In many works on DP graph algorithms, the first step is to design algorithms that require a promise on the degree of the graph for the privacy guarantee. Jain et al.~\cite{JainSW} call such algorithms {\em $D$-restricted}.
\begin{definition}[$D$-restricted differential privacy] Let $V$ be a set of nodes and $k,D,T\in\mathbb{N}$. Let $\Alg$ be a continual release graph algorithm $\allgraphs{V}^T\rightarrow (\mathbb{R}^k)^{T}$. Let $\eps>0$ and $\delta\geq 0$. The algorithm $\Alg$ is $D$-restricted, event-level, $(\eps,\delta)$-edge differentially private (\reddp{D}), if for all event-level edge-neighboring dynamic graph sequences  $\graphseq,\graphseq'\in\allgraphs{V}^T$ of maximum degree $D$ and every $\out\subseteq(\mathbb{R}^k)^T:$ 
\begin{align*}
    \Pr\left[(\Alg(G_1,\dots,G_t)_{t\leq T}\in \out\right]\leq e^{\eps}\Pr\left[(\Alg(G'_1,\dots,G'_t)_{t\leq T}\in \out\right]+\delta.
\end{align*}
The definition of $D$-restricted item-level is analogous.
\end{definition}

\paragraph{$(\alpha,\beta)$-accuracy for graph functions} Let $f:\allgraphs{V}\rightarrow \mathbb{R}^k$. Let $\graphseq=(G_1,\dots,G_T)$ with $G_t\in \allgraphs{V}$ be a dynamic graph sequence. A continual release graph algorithm $\Alg$ is \emph{$(\alpha,\beta)$-accurate} for $f$ on $\graphseq$ if $
    \Pr\left[\max_{t\in[T]}\|\Alg(G_1,\dots,G_t)-f(G_t)\|_{\infty}>\alpha\right]\leq \beta.
$

\begin{definition}[Difference sequence]\label{def:diffseq}
Let $f:\allgraphs{V}\rightarrow \mathbb{R}^k$. Let $\graphseq=(G_1,\dots,G_T)$ 
be a dynamic graph sequence on vertex set $V$ and $G_0=(V,\emptyset)$. The \emph{difference sequence for $f$} is  $(df(\graphseq,t))_{t\in [T]}$, where 
for $t\in[T]$,
$$
df(\graphseq,t)=f(G_t)-f(G_{t-1}).
$$
\end{definition}

\subsection{Problem Definitions}
We consider the following functions on (simple) graphs. We use $\degr{v}$ to denote the degree of a node $v$.
\begin{definition}[Graph functions]\label{def:specific-graph-functions}
    Let $N\in\N$ and  $G=(V,E)$ be a graph with $|V|=N$. 
    \begin{itemize}
        \item {\em \edgecount:} The function $\edgecountmath$ maps a graph $G$ to the number of edges $|E|$.

         \item {\em \trianglecount:} The function $\trianglecountmath$ maps a graph $G$ to the number of triangles in $G$: i.e., 
        $\trianglecountmath((V,E))=|\{ \{x,v,u\}\in \binom{V}{3}: (x,u), (u,v), (v,x) \in E\}|$.
        
        \item {\em $\highdegree{\tau}$:} \sr{For each $\tau\in\N_{\geq 0}$,}
        the function $\highdegreemath{\tau}$ 
        maps a graph $G$ to
        the number of nodes in $G$ of degree at least $\tau$, i.e.,
        $\highdegreemath{\tau}(G)=|\{v\in V: \degr{v}\geq \tau\}|$.

        \item {\em \degreelist:} The function $\degreelistmath$ maps a graph $G$ to the list of node degrees: 
        $\degreelistmath(G)=(\degr{v})_{v\in V}$.
        
        \item {\em \degreehist:} The function $\degreehistmath$ maps 
        a graph $G$ to the histogram $(h_0,h_1,\dots,h_{N-1})$ of node degrees, where $h_j
        =|\{v\in V: \degr{v}=j\}|$ 
        for all $j\in\{0,\dots, N-1\}$.
       
        \item {\em \maxmatch:} The function $\maxmatchmath$ maps a graph $G$ to the size of the maximum matching in $G$.
        \item {\em \conncomp:} The function $\conncompmath$ maps a graph $G$ to the number of connected components in~$G$.
        
    \end{itemize}
\end{definition}

\section{Fully Dynamic Private Triangle Counting}\label{sec:tringle-counting}
This section presents algorithms and lower bounds for triangles counting with event-level privacy and a lower bound for item-level privacy.

\subsection{Lower Bounds for Event-Level \trianglecount via Submatrix Queries}\label{sec:lb-submatrix-queries}
In this section, we prove the following theorem on the lower bounds on the error of fully dynamic, event-level, edge-DP algorithms for \trianglecount.

\begin{theorem}[Lower bounds for event-level \trianglecount]\label{thm:lower_triangle_submatrix}
Let $T,N,D\in\mathbb{N}$ be sufficiently large.
Then every event-level $(0.1,0.02)$-edge-DP algorithm which is $(\alpha,1/6)$-accurate for \trianglecount on all dynamic graph sequences with $N$ nodes and length $T$ satisfies $\alpha=\Omega(\min(T, \sqrt[3]{TN^{2}}, N^2))$.

If the accuracy condition holds instead for all dynamic graph sequences with maximum degree $D$ and length $T$ then $\alpha=\Omega\left(\min\left(T^{1/3}{D^{2/3}},T^{1/4}D\right)\right)$.
\end{theorem}

As discussed in \Cref{sec:techniques}, our reduction relies on submatrix queries, defined next.

\newcommand{\submatrix}[2]{\mathrm{Submatrix}(#1,#2)}
\begin{definition}[Submatrix Query Problem] In the submatrix query problem, the data universe is $\uni=\{0,1\}$. A 
dataset
$Y\in\uni^M$, where $M=n^2$ for some $n\in\N$, can be seen as an $n\times n$ binary matrix, where one data bit corresponds to one entry in the matrix. A submatrix query on $\uni^M$ is defined by two vectors $a,b\in\{0,1\}^n$ and maps $Y\in\uni^M$ to $a^T Y b=\sum_{i=1}^{n}\sum_{j=1}^n a[i]Y[i,j]b[j]$. The problem $\submatrix{n}{k}$ is the problem of answering $k$ submatrix queries of the form $(a,b)$, where $a,b\in\{0,1\}^n$.
\end{definition}

\begin{example}
    Let $Y=(1~0~1~0~0~1~1~1~1)$. 
    We have $M=9$ and $n=3$. To answer a submatrix query $a,b$, where $a=\left(\begin{smallmatrix}1 \\0\\1\end{smallmatrix}\right)$ and $b=\left(\begin{smallmatrix}1\\1\\0\end{smallmatrix}\right)$, we compute $a^TYb=\left(\begin{smallmatrix}1 &0&1\end{smallmatrix}\right)\left(\begin{smallmatrix}1 & 0 & 1\\0&0&1\\1&1&1\end{smallmatrix}\right)\left(\begin{smallmatrix}1\\1\\0\end{smallmatrix}\right)=3$.
\end{example}

\begin{definition}[Accuracy for batch algorithms]  Let $k,n\in\N$. Let $\uni$ be a universe and $f:\uni^n\rightarrow \mathbb{R}^k$ a function. A randomized algorithm $\Alg$ is $(\alpha,\beta)$-accurate for $f$, if for all 
datasets
$y\in\uni^n$ and $z=f(y)$, it outputs $a_1,\dots, a_k$ such that
$\displaystyle\Pr[\max_{i\in[k]}|a_i-z_i|>\alpha]\leq \beta,$ where the probability is taken over the random coin flips of the algorithm.
\end{definition}

The following lower bound is implicitly proved in \cite{EdenLRS23} (Claim 3.5, Lemmas 3.8-3.10).

\begin{lemma}[Submatrix Query lower bound]\label{lem:submatrix}
   There exist constants $c_1 \geq 1$ and $c_2 > 0$ such that, for all large enough $n\in\N$, every $(0.1,0.02)$-differentially private and $(\alpha,1/6)$-accurate algorithm for $\submatrix{n}{c_1 n^2}$  satisfies $\alpha>c_2 n$.
\end{lemma}

\begin{lemma}[Reduction from submatrix queries to \trianglecount]\label{lem:red_triangles_submatrix}
  Let $k,n\in \N$. Let $Y\in\{0,1\}^{n\times n}$ be a 
dataset
and $Q=\big((a^{(m)},b^{(m)})\big)_{m\in[k]}$, where each $a^{(m)},b^{(m)}\in\{0,1\}^{n}$, 
  be a sequence of
$k$ submatrix queries.
 Then for all $w\in \N$,  there exists a transformation from $(Y,Q)$ 
 to a dynamic graph sequence $(G_1,\dots, G_T)$ of $N$-node graphs, where $N=2n+w$ and $T=n^2+4knw$, such that:
 \begin{itemize}
     \item For neighboring 
datasets
    $Y,Y'\in\{0,1\}^{n\times n}$ and a query sequence $Q$, the transformations of $(Y,Q)$ and $(Y',Q)$ give graph sequences which are event-level,  edge-neighboring;
     \item For all $m\in[k]$, 
     the triangle count
     $\trianglecountmath(G_{t_m})=w\cdot (a^{(m)})^T Y  b^{(m)}$, where $t_m=n^2+2nw(2m-1)$. 
 \end{itemize}
 \end{lemma}
 \begin{proof}
 Given a 
dataset
 $Y\in\{0,1\}^{n\times n}$, we construct a graph sequence on $2n+w$ nodes $x_1,\dots, x_n$, $v_1,\dots, v_n$, and $z_1,\dots, z_w$, starting from graph $G_0$ with no edges. We begin with an initialization phase of $n^2$ time steps: For all $i,j\in[n]$, we insert the edge $(x_i,v_j)$ at time $n(i-1)+j$ if $Y[i,j]=1$; otherwise, we do not perform an update at time $n(i-1)+j$.  

 Next, for each $m\in[k]$, we use $4nw$ time steps to process query $(a^{(m)}, b^{(m)})$ as follows. Let $s_m=n^2+4nw(m-1)$ be the time step right before we start processing this query. For all $i\in[n]$, if $a^{(m)}[i]=1$, then, for all $\ell\in[w]$, we insert the edge $(x_i,z_{\ell})$ at time $s_m+(i-1)w+\ell$ 
 and delete it at time $s_m+2nw+(i-1)w+\ell$. Analogously, for all $j\in[n]$, if $b^{(m)}[j]=1$, then, for all $\ell\in[w]$, we insert the edge $(v_j,z_{\ell})$ at time  $s_m+nw+(j-1)w+\ell$ and delete it at time $s_m+3nw+(j-1)w+\ell$. 
 See Figure~\ref{fig:triangle_via_submatrix}. 

 \emph{Correctness.} If $Y$ and $Y'$ differ in coordinate $(i,j)\in[n]\times [n]$, then the corresponding constructed graph sequences differ only in the presence or absence of the initial insertion of edge $(x_i, v_j)$ at time step $n(i-1)+j$.  Thus, they are event-level,  edge-neighboring. For each $m\in[k]$, at time step $t_m=n^2+4nw(m-1)+2nw=s_m+2nw$, the constructed graph has all edges $(x_i,z_{\ell})$ corresponding to the 1-entries of $a^{(m)}$ and all edges $(v_j,z_{\ell})$ corresponding to the 1-entries of $b^{(m)}$. Further, there exists an edge $(x_i,v_j)$ if and only if $Y[i,j]=1$. Thus, the number of triangles at time $t_m$ is equal to $w$ times the number of pairs $(i,j)$ 
 satisfying $a^{(m)}[i]=Y[i,j]=b^{(m)}[j]=1$.
That is, $\trianglecountmath(G_{t_m})=w\cdot a^{(m)}Y  b^{(m)}$.

 \emph{Analysis.} The initialization phase uses $n^2$ time steps.   Processing of each query uses $4nw$ time steps. Thus, $T=n^2+4knw$.
    \begin{figure}[t]
        \centering
        \includegraphics[width=0.35\textwidth]{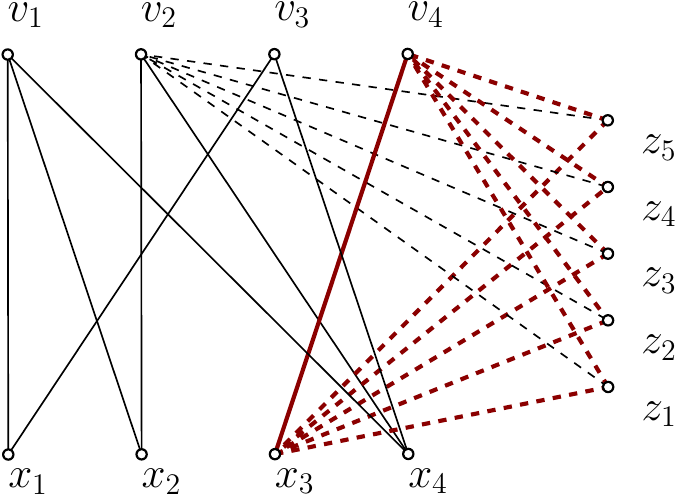}
       \caption{An example of the construction in the proof of \Cref{lem:red_triangles_submatrix} for $n=4$ and $w=5$. The input matrix is $Y=\begin{psmallmatrix} 1& 0 & 1 & 0\\1& 1& 0 & 0\\0&0&0&1\\ 1&1&1&0\end{psmallmatrix}$; the dashed lines show the edges inserted and then deleted for the query $(a,b)$ with $a^T=(0~0~1~0)$ and $b^T=(0~1~0~1)$. There are $5$ triangles between $x_3$, $v_4$, and $z_1,\dots,z_5$, and $(3,4)$ is the only pair $(i,j)$ satisfying $Y[ij]=a[i]=b[j]=1$.}
       \label{fig:triangle_via_submatrix}
    \end{figure}
\end{proof}

\begin{lemma}[Reduction from submatrix queries to \trianglecount, bounded degree]\label{lem:red_triangles_submatrix_boundedD}
  Let $k,n\in \N$. Let $Y\in\{0,1\}^{n\times n}$ be an input 
dataset
and $Q=\big((a^{(m)},b^{(m)})\big)_{m\in[k]}$, where each $a^{(m)},b^{(m)}\in\{0,1\}^{n}$, 
  be a sequence of $k$ submatrix queries.
 Then for all $w,B\in \N$, 
 where $B$ divides $n$,  there exists a transformation from $(Y,Q)$ to a dynamic graph sequence $(G_1,\dots, G_T)$ of $N$-node graphs of maximum degree $D=2B+w$, where $N=\frac{2n^2}{B}+\frac{n^2w}{B^2}$ and $T=n^2+\frac{4kwn^2}{B}$, such that: 
 \begin{itemize}
    \item For neighboring 
datasets
$Y,Y'\in\{0,1\}^{n\times n}$ and a query sequence $Q$, the transformations of $(Y,Q)$ and $(Y',Q)$ give graph sequences which are event-level,  edge-neighboring;
     \item For all $m\in[k]$, 
     the triangle count
     $\trianglecountmath(G_{{t_m}})=w\cdot a^{(m)}Y  b^{(m)}$, where $t_m=n^2+(m-1)\frac{4n^2w}{B}+\frac{2n^2w}{B}$.
 \end{itemize}
 \end{lemma}
\begin{proof}

    \begin{figure}[t]
        \centering
        \includegraphics[width=0.7\textwidth]{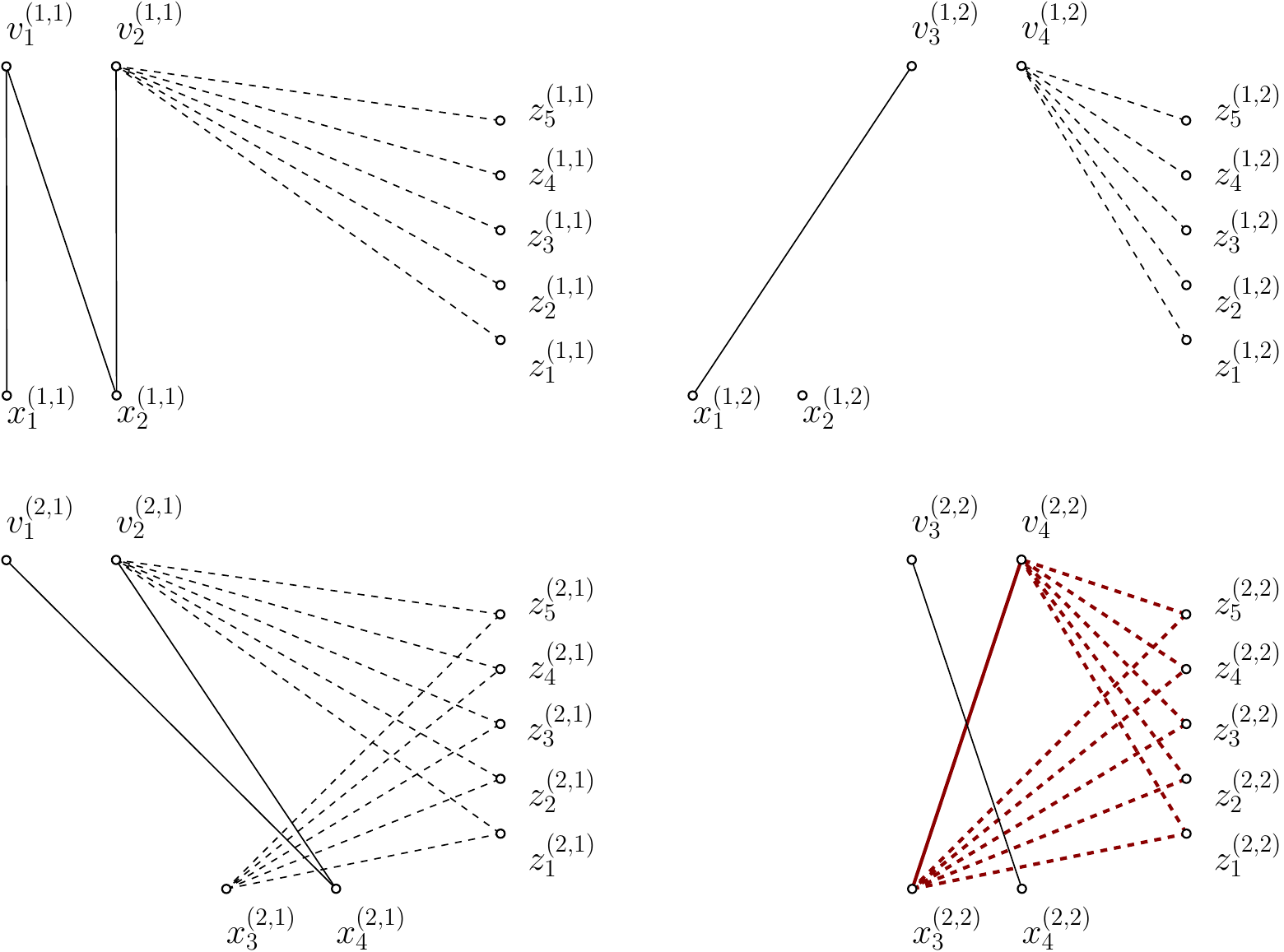}
       \caption{An example of the construction in the proof of \Cref{lem:red_triangles_submatrix_boundedD} for $n=4$, $B=2$ and $w=5$. The input matrix is $Y=\begin{psmallmatrix} 1& 0 & 1 & 0\\1& 1& 0 & 0\\0&0&0&1\\ 1&1&1&0\end{psmallmatrix}$; the dashed lines show the edges inserted and then deleted for the query $(a,b)$ with $a^T=(0~0~1~0)$ and $b^T=(0~1~0~1)$.} 
       \label{fig:triangle_via_submatrix_many}
    \end{figure}
  Given a 
dataset
$Y\in\{0,1\}^{n\times n}$ and $B,w\in\mathbb{N}$, we construct a graph sequence on $\frac{n^2}{B^2}\times (2B+w)$ nodes, starting from graph $G_0$ with no edges.  At a high level, we divide the matrix $Y$ into submatrices of size $B\times B$, and for each submatrix, we build a graph as in \Cref{lem:red_triangles_submatrix}. 
  Specifically, for all $p_1,p_2\in\left[\frac{n}{B}\right]$, define $Y^{(p_1,p_2)}$ as the $B\times B$ matrix with $Y^{(p_1,p_2)}[i,j]=Y[(p_1-1)B+i,(p_2-1)B+j]$ for all $i,j\in[B]$. We create a {\em gadget} for each $(p_1,p_2)\in \left[\frac{n}{B}\right]^2$ by allocating $2B+w$ nodes for it. In the initialization phase, each gadget is updated by running the initialization phase from the proof of \Cref{lem:red_triangles_submatrix} for $Y^{(p_1,p_2)}$ and $w$.
  The initialization phase uses $\left(\frac{n}{B}\right)^2B^2=n^2$ time steps.
    
    Then, for each $m\in[k]$, to process query $(a^{(m)},b^{(m)})$, we run the processing procedure from the proof of \Cref{lem:red_triangles_submatrix} for every $p_1,p_2\in\left[\frac{n}{B}\right]$ for $(a^{(m)}[(p_1-1)B+1,p_1B],b^{(m)}[(p_2-1)B+1,p_2B])$ within the gadget for $(p_1,p_2)$. That is, for all $p_1\in\left[\frac{n}{B}\right]$, we use $Bw\frac{n}{B}$ time steps to insert the edges corresponding to $a^{(m)}[(p_1-1)B+1,p_1B]$ into all gadgets for $(p_1,p_2)$ for all $p_2\in\left[\frac{n}{B}\right]$. Similarly, for all $p_2\in\left[\frac{n}{B}\right]$, we insert the edges corresponding to $b^{(m)}[(p_2-1)B+1,p_2B])$ into all gadgets for $(p_1,p_2)$ for all $p_1\in \left[\frac{n}{B}\right]$. This uses in total $2\frac{n^2 w}{B}$ time steps. Afterwards, we use $2\frac{n^2 w}{B}$ time steps to delete the edges corresponding to the $m$th query for all gadgets (as in the proof of \Cref{lem:red_triangles_submatrix}).
    
    \emph{Correctness.} For every pair $(i,j)\in[n]^2$, there exists exactly one pair $(p_1,p_2)$ such that $i\in [(p_1-1)B+1,p_1B]$ and $j\in[(p_2-1)B+1,p_2B]$. Thus, if $Y$ and $Y'$ differ in entry $(i,j)$, then only the gadget for $(p_1,p_2)$ will differ in the initial insertion of the corresponding edge. 
    We finish inserting all edges corresponding to query $(a^{(m)},b^{(m)})$ at time step $t_m=n^2+(m-1)\frac{4n^2w}{B}+{2n^2w}{B}$.
    For each pair $(p_1,p_2)\in\left[\frac{n}{B}\right]\times \left[\frac{n}{B}\right]$, the number of triangles in the gadget for $(p_1,p_2)$ at time $t_m$ is  $wa^{(m)}[(p_1-1)B+1,p_1B] Y^{(p_1,p_2)} b^{(m)}[(p_2-1)B+1,p_2B])$, and $\sum_{p_1,p_2\in\left[\frac{n}{B}\right]}wa^{(m)}[(p_1-1)B+1,p_1B] Y^{(p_1,p_2)} b^{(m)}[(p_2-1)B+1,p_2B])=wa^{(m)}Y^{(p_1,p_2)}b^{(m)}$.
    
    \emph{Analysis.} The total number of nodes is  $\frac{n^2}{B^2}\times (2B+w)=\frac{2n^2}{B}+\frac{n^2w}{B^2}$. The degree of each $G_t$ is at most $2B+w$. Initialization takes $n^2$ time steps, and each query uses $4\cdot \frac {n^2} B \cdot w$ time steps. Thus, $T=n^2+4k\cdot\frac {n^2} B\cdot w$.
\end{proof}

\begin{proof}[Proof of Theorem~\ref{thm:lower_triangle_submatrix}]
     Let $c_1$ be as in \Cref{lem:submatrix}. First, we prove the lower bound in terms of $T$ and $N$. Set $w=\left\lfloor\min\left(\frac N 3,\frac{T-1}{4}\right)\right\rfloor$ and $n=\left\lfloor\min\Big(\frac N 3, \big(\frac{T}{4wc_1+1}\big)^{1/3}\Big)\right\rfloor$. For answering $k=c_1n^2$ queries, the construction in \Cref{lem:red_triangles_submatrix} has $2n+w\leq N$ nodes and $n^2+4c_1n^3 w\leq (4c_1w+1)n^3\leq T$ time steps. We add $N-(2n+w)$ dummy nodes and $T-(n^2+4c_1n^3w)$ time steps without updates in order to get a dynamic graph sequence of $N$ nodes and $T$ time steps. Let $\Alg$ be an event-level $(0.1,0.02)$-edge DP algorithm for \trianglecount which is $(\alpha,\frac 1 6)$-accurate for this graph sequence. Then \Cref{lem:red_triangles_submatrix} gives an algorithm which, with probability at least $\frac 56$, has additive error at most $\frac{\alpha}{w}$ on all submatrix queries. Thus, by \Cref{lem:submatrix}, we have $\frac{\alpha}{w}>c_2 n$ and therefore $\alpha> c_2nw$, where $c_2$ is as in \Cref{lem:submatrix}.  Consider three cases:
    \begin{enumerate}
        \item $T\leq N$: Then $w=\Theta(T)$ and $n=\Theta(1)$, and the lower bound is $\Omega(nw)=\Omega(T)$.
        
        \item $T\in(N,N^4$): Then $w=\Theta(N)$ and $n=\Theta\left(\frac{T}{N}\right)^{1/3}$, and the lower bound is $\Omega(nw)=\Omega(T^{1/3}N^{2/3})$.

        \item $T\geq N^4$
        : Then $N\leq T$ and $N\leq \left(\frac T N\right)^{1/3}$, so  $w=n=\Theta(N)$, and the lower bound is $\Omega(nw)=\Omega(N^2)$.
        \end{enumerate}
        In all three cases, the lower bound is $\Omega(nw)=\Omega(\min(T, T^{1/3}N^{2/3}, N^2))$.

Now we prove the lower bound in terms of $T$ and $D$. Set $w=\lfloor D/3\rfloor$ and  $n'=\left\lfloor\min\left(\sqrt[4]{\frac{T}{5c_1}},
\sqrt[3]{\frac{3T}{5Dc_1}}
\right)\right\rfloor$. If $n'\leq \lfloor D/3\rfloor$, let $n=n'$, else, let $n$ be an integer in $[n'-\lfloor D/3\rfloor,n']$ such that $\lfloor D/3\rfloor$ divides $n$. Let $B=\min(n,\lfloor D/3\rfloor)$.  For answering $k=c_1n^2$  queries, the construction from \Cref{lem:red_triangles_submatrix_boundedD} gives a graphs sequence of maximum degree $D$ and length  $n^2+\frac{4c_1n^4w}{B}\leq \frac{5c_1n^4w}{B}\leq T$. We add extra dummy time steps  to get a graph sequence of length $T$. Let $\Alg$ be an event-level $(0.1,0.02)$-edge DP algorithm for \trianglecount which is $(\alpha, 1/6)$-accurate for this graph sequence. Then \Cref{lem:red_triangles_submatrix_boundedD} gives a $(0.1,0.02)$-DP algorithm with additive error at most $\frac{\alpha}{w}$ on all submatrix queries. Then \Cref{lem:submatrix} implies  $\alpha>c_2wn$, where $c_2$ is as in \Cref{lem:submatrix}. Thus, $\alpha=\Omega\left(\min\left(T^{1/4}D, T^{1/3}{D^{2/3}}\right)\right)$.
\end{proof}

\subsection{Event-Level Private Algorithm for Triangle Count}
In this section, we present our fully dynamic event-level, edge-DP algorithm for triangle counting. Our algorithmic results on this problem are summarized in the following theorem.

\begin{theorem}[Upper bound for event-level \trianglecount]\label{thm:upper_triangle} 
For all $\eps,\delta, \beta\in(0,1)$ and $T,N\in\N$
    there exists an event-level \ededgedp algorithm which is $(\alpha,\beta)$-accurate for \trianglecount on all dynamic graph sequences with $N$ nodes and length $T$, where $\alpha=O\left(\min\left(\frac{\sqrt{TN}}\eps(\ln\frac 1 \delta)\log^{3/2}\frac T \beta,N\left(\frac T{\eps^2}(\ln \frac 1 \delta)\ln \frac T \beta\right)^{1/3}, N^3\right)\right)$. Further, for all dynamic graph sequences with degree at most $D$, for $D\in\N$ and $D\leq N$, the algorithm satisfies $\alpha=O\left(\min\left(\sqrt{TD},DT^{1/3}, ND^2\right)\cdot\mathrm{poly}\left(\frac 1\eps\log \frac{TN}{\beta\delta}\right)\right)$.

\end{theorem}

The minimum expressions in the bounds in \Cref{thm:upper_triangle} are the result of selecting the best algorithm for each setting of parameters. Later (in \Cref{cor:upperbound_triangle_item-level}), we will present item-level (and thus also event-level) algorithms for triangle counting.
The algorithm in this section, whose performance is summarized in \Cref{lem:ed_upper_general},  has better error than the item-level
edge-DP algorithm for the same problem in \Cref{cor:upperbound_triangle_item-level} when the 
number of nodes is large: specifically, $N\geq T^{1/3}$.

\begin{theorem}[Event-level DP triangle count]\label{lem:ed_upper_general}
    For all $\eps,\delta, \beta\in(0,1)$ and $T,N, D\in\N$, there exists an event-level \ededgedp algorithm which is $(\alpha,\beta)$-accurate for \trianglecount on all dynamic graph sequences with $N$ nodes, maximum degree $D$ and length $T$, where $\alpha=O(\sqrt{TD}\cdot\mathrm{poly}(\frac 1\eps\log \frac{TN}{\beta\delta}))$.
\end{theorem}
Theorem~\ref{lem:ed_upper_general} follows from \Cref{lem:ed_upper_2} that gives a $D$-restricted version of the algorithm and Theorem~\ref{thm:d-restricted} that gives a transformation from $D$-restricted to general algorithms.
The $D$-restricted version of the algorithm is based on running a binary tree mechanism on the difference sequence, combined with a careful analysis of the $L_2$-sensitivity of the partial sums used by the mechanism.

\begin{lemma}[$D$-restricted private triangle count]\label{lem:ed_upper_2}Let $\eps,\delta, \beta\in(0,1)$ and $T, D\in\N$. 
    There exists an event-level \reddp{D} algorithm which is $(\alpha,\beta)$-accurate for \trianglecount on all dynamic graph sequences 
    of maximum degree $D$ and length $T$, where $\alpha=O\left(\frac 1\eps\sqrt{TD\ln\frac 1 \delta}\log^{3/2}\frac T \beta\right)$.
\end{lemma}

\begin{proof}
Let $\graphseq=(G_t)_{t\in[T]}$ be a dynamic graph sequence of degree at most $D$. Recall that $\trianglecountmath(G)$ returns the number of triangles in graph $G$.
Let $(d\trianglecountmath(\graphseq,t))_{t\in[T]}$ be the difference sequence for $\trianglecountmath$. (See \Cref{def:diffseq}).

    Our general strategy is to run a counting mechanism similar to the binary tree mechanism by Dwork~et~al. \cite{DworkNPR10} to compute a running estimate on 
    $\trianglecountmath(G_t)=\sum_{t'\in [t]}d\trianglecountmath(\graphseq,t').$
    However, we use Gaussian noise to compute the noisy counts for nodes in the binary tree. We show that when building the binary tree over $d\trianglecountmath$, the $L_2$-sensitivity of the vector of the counts of the binary tree nodes is 
    $O(\sqrt{TD\log T})$. We next describe this approach in detail. 

 Let  $\mathcal{I}_{\ell}=\{[j\cdot 2^{\ell}+1, (j+1)2^{\ell}],0\leq j\leq \lceil T/2^{\ell}\rceil -1\}$ for all (integer) levels $\ell\in[0,\log T]$ and $\mathcal{I}=\bigcup_{0\leq \ell \leq \lfloor \log T \rfloor}\mathcal{I}_{\ell}$. For an interval $[a,b]$ in $\mathcal{I}$, define $s_{[a,b]}(\graphseq)=\sum_{a\leq t\leq b}d\trianglecountmath(\graphseq,t)$. 
 We first obtain a private estimate of $s_{[a,b]}$ for all intervals $[a,b]\in \mathcal{I}$, by computing $\tilde{s}_{[a,b]}=s_{[a,b]}+Y_{[a,b]}$, where $Y_{[a,b]}$ is independently drawn from $N(0,\sigma^2)$ with $\sigma=\frac{12}{\eps}\sqrt{TD\left(\ln\frac{2}{\delta}\right)\log T}$. 
 For any interval $[1,t]$, there exists a set $I_t\subseteq \mathcal{I}$ such that (i) $\bigcup_{[a,b]\in I_t}[a,b]=[1,t]$, (ii) $\bigcap_{[a,b]\in I_t}=\emptyset$, and (iii) $|I_t|\leq \lfloor \log T \rfloor +1$. We compute an estimate of $\trianglecountmath(G_t)$ as $\tilde{f}_{\Delta}(G_t)=\sum_{[a,b]\in I_t}\tilde{s}_{[a,b]}=\sum_{[a,b]\in I_t}s_{[a,b]}+Y_{[a,b]}$.  

\begin{claim}\label{claim:uppertriangle} The $L_2$-sensitivity of $(s_{[a,b]})_{[a,b]\in \mathcal{I}}$ is \sr{at most} $6\sqrt{TD\log T}$. 
\end{claim} 
\begin{proof}
Let $\graphseq=(G_t)_{t\in[T]}$ and $\graphseq'=(G'_1,\dots, G'_T)$ be two event-level, edge-neighboring dynamic graph sequences of degree at most $D$. Let $t_1$ be the first time step where $\graphseq$ and $\graphseq'$ differ. 
W.l.o.g., 
suppose an edge  $e^*=(u^{*},v^{*})$ is inserted into $G_{t_1}$, but  $G'_{t_1}=G'_{t_1-1}$.
Let $t_2$ be the time step where $e^*$ is deleted from $\graphseq$ but not $\graphseq'$ (set $t_2=T+1$ 
if $e^*$ is not deleted in $\graphseq$ after $t_1$). Then $|d\trianglecountmath(\graphseq,t_1)-d\trianglecountmath(\graphseq',t_1)|\leq D-1$ and, if $t_2\leq T$ then $|d\trianglecountmath(\graphseq,t_2)-d\trianglecountmath(\graphseq',t_2)|\leq D-1$, since $e^*$ can be involved in at most $D-1$ triangles. For all $t\in(t_1,t_2)$, if edge $e=(u,v)$ is inserted at step $t$, it can form at most one triangle with $e^{*}$. Thus, $|d\trianglecountmath(\graphseq,t)-d\trianglecountmath(\graphseq',t)|\leq 1$ 
for all $t\notin\{t_1,t_2\}$. 

Now, fix a level $\ell\in\{0,\dots, \lfloor \log T \rfloor \}$. For each interval $[a,b]\in \mathcal{I}_{\ell}$ (which has length $2^{\ell}$), we have $|s_{[a,b]}(\graphseq)-s_{[a,b]}(\graphseq')|\leq D+2^{\ell}$ 
if $t_1\in[a,b]$ or $t_2\in[a,b]$, and $|s_{[a,b]}(\graphseq)-s_{[a,b]}(\graphseq')|\leq 2^{\ell}$, otherwise.
   
     For all $t\in [T]$, the total triangle count can differ by at most $D-1$ for $G_t$ and $G'_t$, that is, $|\trianglecountmath(G_t)-\trianglecountmath(G'_t)|\leq D-1$. 
     For each interval $[a,b]\in\mathcal{I}$, we have $s_{[a,b]}(\graphseq)=\trianglecountmath(G_b)-\trianglecountmath(G_{a-1})$
     and
     $|s_{[a,b]}(\graphseq)-s_{[a,b]}(\graphseq')|=
     |\trianglecountmath(G_b)-\trianglecountmath(G_{a-1})-\trianglecountmath(G'_b)+\trianglecountmath(G'_{a-1})|\leq |\trianglecountmath(G_b)-\trianglecountmath(G'_b)|+|\trianglecountmath(G'_{a-1})-\trianglecountmath(G_{a-1})|
     \leq 2D$. 
  
    There are at most two intervals $I\in\mathcal{I}_{\ell}$ such that $t_1\in I$ or $t_2\in I$. For such an interval $I$, we have $|s_I(\graphseq)-s_I(\graphseq')|\leq D+\min(D,2^{\ell})$. For all other intervals  $[a,b]\in \mathcal{I}_{\ell}$, we have $|s_{[a,b]}(\graphseq)-s_{[a,b]}(\graphseq')|\leq \min(2D,2^{\ell})$. Since $\mathcal{I}_{\ell}$ contains at most $T/2^{\ell}$ intervals, the $L_2$-sensitivity of $(s_{[a,b]})_{[a,b]\in \mathcal{I}_{\ell}}$ is at most 
    
    $2D+\sqrt{\frac{T}{2^{\ell}}}\min(2D,2^{\ell})
    \leq 2D+\sqrt{\frac{T}{\min(2D,2^{\ell})}}\min(2D,2^{\ell})
    \leq 2D+\sqrt{T \cdot \min(2D,2^{\ell})}
    \leq 2D+ \sqrt{2TD},$
    
  \noindent  which is at most $4\sqrt{TD}$, since $D\leq T$. Thus, the $L_2$-sensitivity of $(s_{[a,b]})_{[a,b]\in \mathcal{I}}$ is at most 
  $\sqrt{\sum_{\ell=0}^{\lfloor \log T \rfloor}16TD}\leq 6\sqrt{TD\log T}.$
    \end{proof}
  By \Cref{lem:gaussianmech}, the algorithm that returns
 $\tilde{s}_{[a,b]}=s_{[a,b]}+Y_{[a,b]}$, where every $Y_{[a,b]}$ is independently drawn from $N(0,\sigma^2)$, is $(\eps,\delta)$-DP. 
   Recall that $\tilde{f}_{\Delta}(G_t)=\sum_{[a,b]\in I_t}\tilde{s}_{[a,b]}=\sum_{[a,b]\in I_t}s_{[a,b]}+Y_{[a,b]}$ and note that $\sum_{[a,b]\in I_t}s_{[a,b]}=\trianglecountmath(G_t)$.
   The random variable $Z_t=\sum_{[a,b]\in I_t}Y_{[a,b]}$ has distribution $N(0,\hat{\sigma}^2)$, where $\hat{\sigma}^2\leq (\lfloor \log T \rfloor +1)\sigma^2$. By \Cref{lem:gaussian_tail}, 
   for all $\beta\in(0,1)$, we have that $\Pr[|Z_t|\geq \hat{\sigma}\sqrt{2\ln(T/\beta)}]\leq \beta/T$. By a union bound over the $T$ time steps, with probability at least $1-\beta$, estimates returned at all steps have error at most $O(\hat\sigma\sqrt{\log(T/\beta)})$.
   Thus, the algorithm is $(\alpha,\beta)$-accurate, for 
   $$\alpha=O\left(\frac 1 \eps \sqrt{TD\left(\ln\frac 1 \delta\right)\log T}\sqrt{\log T}\sqrt{\log(T/\beta)}\right)=O\left(\frac 1\eps\sqrt{TD\ln\frac 1 \delta}\log^{3/2}\frac T \beta\right).  $$
\end{proof}

\subsection{Lower Bound for Item-Level \trianglecount}
In this section, we prove the following theorem that gives a lower bound on triangle counting with item-level edge-DP.

\begin{theorem}[Item-level lower bound for triangle count]\label{thm:lb-triangle_item-level}
   For all $\eps\in(0,1]$, $\delta\in[0,1)$, and sufficiently large $T, N\in\N$, every item-level, $(\eps,\delta)$-edge DP algorithm for $\trianglecountmath$ which is $(\alpha,\frac{1}{3})$-accurate on all dynamic graph sequences with $N$ nodes and length $T$ satisfies 
   \begin{enumerate}
       \item if $\delta>0$ and $\delta=o(\frac 1 T)$, then $\alpha=\Omega\left(\min\left(\frac{T^{7/6}}{(\eps\log T)^{1/3}},\frac{NT^{1/3}}{\eps^{2/3}\log^{2/3} T},N^3\right)\right)$.
       \item if $\delta=0$, then $\alpha=\Omega\left(\min\left(\frac{T^{5/4}}{\eps^{1/4}},N\sqrt{\frac T \eps },N^3\right)\right)$.
   \end{enumerate}
\end{theorem}

To prove the theorem, we give a reduction from the \emph{1-way marginals problem} in the batch model, defined next.
\begin{definition}[1-way marginals]\label{def:marginals}
 Let $d,n\in \N$ and let $\uni=\{0,1\}^d$. In the $\marginals{n}{d}$ problem, the input is a 
dataset
 $Y=(Y_1,\dots,Y_n)\in\uni^n$, and the goal is to compute the vector $(q_1(Y),\dots,q_d(Y))$, where $q_j(Y)=\frac{1}{n}\sum_{i=1}^{n} Y_{i}[j]$ for all $j\in[d]$. 

 A randomized algorithm $\Alg$ is $(\alpha,\beta)$-accurate for $\marginals{n}{d}$, if for all 
datasets
 $Y\in\{\{0,1\}^d\}^n$, it outputs $a_1,\dots, a_d$ such that
$$\Pr[\max_{j\in[d]}|a_j-q_j(Y)|>\alpha]\leq \beta,$$ where 
the probability is taken over the random coin flips of the algorithm.
\end{definition}

The following lower bound is taken from \cite{JainRSS23}, which in turn is based on \cite{BunUV18,HardtT10}.
\begin{lemma}[Lower bound for Marginals \cite{BunUV18,HardtT10,JainRSS23}]\label{lem:marginals}
    For all $\eps\in(0,1]$, $\delta\in[0,1)$, $\alpha\in(0,1)$, and $d,n\in\mathbb{N}$, every $(\eps,\delta)$-DP algorithm that is  $\left(\alpha, \frac{1}{3}\right)$-accurate for $\marginals{n}{d}$ satisfies:
    \begin{itemize}
        \item if $\delta>0$ and $\delta=o\left(\frac 1 n \right)$, then $\alpha=\Omega\left(\frac{\sqrt{d}}{n\eps\log d}\right)$.
        \item if $\delta=0$, then $\alpha=\Omega\left(\frac{d}{n\eps}\right)$.
    \end{itemize}
\end{lemma}
\begin{lemma}[Reduction from Marginals to \trianglecount]\label{lem:red:triangle-count}
    Let $n,d\in\N$ and $Y\in\{\{0,1\}^d\}^n$. Then for all $w\in \N$, 
    there exists a transformation from \sr{$Y$} 
    to a dynamic graph sequence $(G_1,\dots, G_T)$ of $N$-node graphs, where $T= 2\lceil \sqrt{n}\rceil w+2nd$ and $N=2\lceil \sqrt{n}\rceil +w $, such that
    \begin{itemize}
        \item The transformations of neighboring $Y,Y'\in\{\{0,1\}^d\}^n$ give item-level,  edge-neighboring graph sequences;
        \item Let $t_0=2\lceil \sqrt{n}\rceil w$. For all $j\in[d]$ and 
        $t_j=t_0+(2j-1)n,$
        we have $\trianglecountmath(G_{t_j})=w\sum_{i=1}^nY_i[j]$.
    \end{itemize}
\end{lemma}
\begin{proof}
We define a set $V=V_0\cup V_1\cup W$, where $V_0, V_1$ and $W$ are pairwise disjoint, $|V_0|=|V_1|=\lceil\sqrt{n}\rceil,$ and $|W|=w$. There are at least $n$ node pairs $(v_0,v_1)\in V_0\times V_1$. We give them an arbitrary order $e_1,\dots, e_{\lceil\sqrt{n}\rceil^2}$. In particular, there exist the node pairs $e_1,\dots, e_n$.

In the initialization phase, we insert edge $(v,u)$ for all $v\in V_0\cup V_1$ and all $u\in W$, in an arbitrary but fixed order. This takes $2\lceil \sqrt{n}\rceil w$ 
time steps. 
Then, for all $j\in[d]$ and $i\in[n]$, if $Y_i[j]=1$ then we insert $e_i$ at time $t_0+2(j-1)n+i$ \sr{and delete $e_i$ at time $t_0+(2j-1)n+i$; otherwise, we do nothing at these time steps. This takes $2nd$ time steps, and $T= 2\lceil \sqrt{n}\rceil w+2nd$ for the whole construction. }

  \emph{Correctness.} If $Y$ and $Y'$ are neighboring input 
    datasets
differing in row $i^*$, then the resulting graph sequences for $Y$ and $Y'$ differ only in insertions and deletions related to $e_{i^*}$. Thus, they are item-level, edge-neighboring. The only triangles that can form have to include an edge from $W$ to $V_0$, from $V_0$ to $V_1$, and from $V_1$ to $W$. Let $G_t=(V, E_t)$. Since every node in $W$ is connected to every node $V_0\cup V_1$, the number of triangles is exactly $w|\{e_i:e_i\in E_t\}|$. At time step $t_j=t_0+(2j-1)n$, the graph $G_{t_j}$ includes an edge $e_i$ if and only if $Y_i[j]=1$. Thus, the total triangle count is equal to $w\sum_{i=1}^nY_i[j]$.
%
%
\end{proof}

    \begin{proof}[Proof of \Cref{thm:lb-triangle_item-level}]

   {\bf Case $\delta>0$:} Given $T$ and $N$, we set $d=\big\lfloor\left(T\eps\log({T\eps)}\right)^{2/3}\big\rfloor$, and $n=\lfloor\min\left(\frac{\sqrt{d}}{6\eps\log d}, \frac{N^2}{\sr{16}}\right)\rfloor$ and $w=\big\lfloor\min\left(\frac{N}{2}, \frac{T}{4\lceil\sqrt{n}\rceil}\right)\big\rfloor$.  The reduction from \Cref{lem:red:triangle-count} 
   gives a sequence with at most $N$ nodes and $T/2 + 2nd$ updates, where 
   $$2nd=\frac{d^{3/2}}{3\eps\log d}\leq\frac{{T\eps}\log ({T\eps})}{3\eps\cdot\frac{2}{3}\log({T\eps}\log ({T\eps}))}\leq \frac{T}{2}.$$
   We can add  dummy nodes and dummy time steps to get a dynamic $N$-node graph sequence of length $T$. If we can estimate $\trianglecountmath(G_t)$ with error up to $\alpha$ at all time steps with probability at least $2/3$, then for all $j\in[d]$, we can estimate the $j$th marginal given by $\frac{1}{n}\sum_{i\in n}Y_i[j]=\frac{1}{wn}\trianglecountmath(G_{t_j})$ up to error $\frac{2\alpha}{nw}$.

    \Cref{lem:marginals}
    gives $\alpha=\Omega\left(nw\cdot\min\left(\frac{\sqrt{d}}{n\eps\log d},1\right)\right)=\Omega(nw)=\Omega\left(\min\left(\frac{T^{7/6}}{(\eps\log T)^{1/3}},\frac{NT^{1/3}}{\eps^{2/3}\log^{2/3} T},N^3\right)\right)$.

    {\bf Case $\delta=0$:} Given  $T$ and $N$, set $d=\lfloor\sqrt{T\eps}\rfloor$ and $n=\lfloor\min\left(\frac{d}{2\eps},\frac{N^2}{4}\right)\rfloor$ and $w=\big\lfloor\min\left(\frac{N}{2}, \frac{T}{4\lfloor\sqrt{n}\rfloor}\right)\big\rfloor$. The reduction from \Cref{lem:red:triangle-count} gives a sequence with at most $N$ nodes and at most $T$ updates, and we can add dummy nodes and dummy time steps to get a dynamic $N$-node graph sequence of length $T$. By \Cref{lem:marginals}, we have 
    $\alpha=\Omega\left(nw\cdot\min\left(\frac{d}{n\eps},1\right)\right)=\Omega(nw)=\Omega\left(\min\left(\frac{T^{5/4}}{\eps^{1/4}},N\sqrt{\frac T \eps },N^3\right)\right)$.
    \end{proof}

\section{General Algorithmic Tools for DP Fully Dynamic Algorithms}\label{sec:algorithmic-tools}
This section presents general algorithmic tools that can be used for all graph problems we discussed.
\subsection{Transformation from Degree-Restricted to Private}\label{sec:transformation-from-degree-restricted-to-private}
We start with our general transformation in the event-level setting from a $D$-restricted edge-DP algorithm with an error bound that depends on $D$ to an edge-DP algorithm for {\em all} dynamic graph sequences and has an error bound depending on the maximum degree of the sequence.
The error bounds are the same as for the $D$-restricted case, up to factors in $\frac 1\eps\log(\frac {TN}{\beta\delta})$.

\begin{theorem}\label{thm:d-restricted}
   Let $T,N,k\in \N$ and $f$ be a function $\allgraphs{V}\rightarrow \mathbb{R}^k$. Let $\alpha:\N\times(0,1)\rightarrow \R^+$ be a function of the degree $D$ and the error probability $\beta$, such that $\alpha(D,\beta)$ is  nondecreasing in $D$  and nonincreasing in $\beta$. If for all $\beta\in(0,1)$, $\eps>0$, $D\in\mathbb{N}$, and $\delta\in[0,1)$, there is an event-level \reddp{D} algorithm $\Alg_D$ which is $(\alpha(D,\beta),\beta)$-accurate for $f$ on all dynamic graph sequences with $N$ nodes, maximum degree $D$ and length $T$, 
    then for all $D'\in \N$, $\eps'>0$, $\delta'\in(0,1)$ and 
    $\beta'\in(0,1)$, there exists
     an 
    event-level $(\eps',\delta')$-edge DP algorithm for $f$  which is $\left(\alpha\left(D'',\min\left(\frac{\delta'}{2+2e^{\eps'}},\frac{\beta'}{2+\log N}\right)\right),\beta'\right)$-accurate on all dynamic graph sequences with $N$ nodes, maximum degree $D'$, and length $T$, where $D''=O(D'+\frac{\log T\cdot \log N}{\eps'} \log\frac{TN\log N}{\beta'\delta'})$. The value $D'$ does not need to be given to the algorithm.
\end{theorem}

As one of the main tools, we first give an algorithm for accurately estimating the degree list of a dynamic graph sequence. It is based on computing a running sum of the difference sequence (see Definition~\ref{def:diffseq}), i.e., the change of the function value between two time steps, via known counting mechanisms. A similar strategy was used in \cite{FichtenbergerHO21} for other graph functions.
\begin{lemma}[Algorithm for \degreelist] \label{lem:estimating_deg}
For all $T,N,D\in \N$, $\eps>0$, and $\beta\in(0,1)$, there exists an event-level \eedgedp algorithm for \degreelist\ which is $(\alpha,\beta)$-accurate on all dynamic $N$-node graph sequences of length $T$,
where $\alpha=O\left(\frac{\log T}\eps \log\frac{TN}{\beta}\right)$.
\end{lemma}
\begin{proof} Let $\graphseq=(G_1,\dots, G_T)$ be a dynamic graph sequence on a node set $V$ with $|V|=N$.
  For every node $v\in V$, we compute an event-level edge-DP estimate of its degree using a continual counting algorithm.  For $t\in[T]$, let $f_v(G)=\deg_G(v)$, where $\deg_G(v)$ denotes the degree of $v$ in $G$. We use a continual counting algorithm on the difference sequence $(df_v(\graphseq,t))_{t\in[T]}$ (recall Definition~\ref{def:diffseq}).
  We have that $df_v(\graphseq,t)\in\{-1,0,1\}$ 
  for all $v\in V$ and all $t\in[T]$. Additionally, for each $t\in[T]$, the vector $(df_v(\graphseq,t))_{v\in V}$ has at most $2$ 
  nonzero entries, corresponding to the endpoints of the edge updated at time step $t$.
  Further, for all neighboring sequences $\graphseq$ and $\graphseq'$, the resulting difference sequences differ in at most two time steps.
    Using \Cref{lem:dphist}, Item~\ref{lem:dphist_eps}, the lemma follows. 
    \end{proof}
Next, we prove \ts{Lemma~\ref{lem:d-rest}, which is the key lemma} of this section. The main idea is to keep a running estimate of the maximum degree, and run the corresponding degree restricted mechanism, until the maximum degree increases significantly. We then re-initialize a new restricted mechanism with a higher degree bound.  To get Theorem~\ref{thm:d-restricted} from \Cref{lem:d-rest}, we plug in the parameters to get an event-level $(\eps',\delta')$-edge-DP algorithm and an error bound that holds with probability at least $1-\beta'$, where $\eps'>0$, $\delta'\in(0,1)$ and $\beta'\in(0,1)$ can be chosen freely. 

In the proof of the following lemma, we use the notion of $(\eps,\delta)$-indistinguishability:
\begin{definition}[$(\eps,\delta)$-indistinguishability]
Let $\uni$ and $\mathcal{Y}$ be two sets. 
Random variables $Y$ and $Y'$, mapping $\uni$ to $\cY$, are \emph{$(\eps,\delta)$-indistinguishable} if for all $\out\subseteq \mathcal{Y}$,
\begin{align*}
    \Pr[Y\in \out]&\leq e^{\eps}\Pr[Y'\in \out]+\delta;\\
    \Pr[Y'\in \out]&\leq e^{\eps}\Pr[Y\in \out]+\delta.
\end{align*}
We denote that $Y$ and $Y'$ are $(\eps,\delta)$-indistinguishable by $Y\approx_{\eps,\delta}Y'$.
\end{definition}
That is, an algorithm $\Alg$ is $(\eps,\delta)$-DP if and only if $\Alg(x)\approx_{\eps,\delta}\Alg(y)$
for all neighboring $x$ and $y$.
\begin{lemma}[Transformation from degree-restricted privacy]\label{lem:d-rest}
Let $T,N,k\in \N$ and let $f$ be a function $\allgraphs{V}\rightarrow \mathbb{R}^k$. Let $\alpha:\N\rightarrow \R^+$ be a nondecreasing function of the degree $D$. Let $\eps>0$ and $\delta\in[0,1)$ and $\beta,\beta_s\in(0,1)$.  Assume that for every $D\in \N$, there is an event-level \reddp{D} algorithm $\Alg_D$ which is $(\alpha(D),\beta)$-accurate for $f$ on all dynamic graph sequences with $N$ nodes, maximum degree $D$ and length $T$. 
    Then 
    there exists
     an event-level 
     $(\eps(2+\log N),\delta(1+\log N)+\beta_s(1+e^{\eps}))$-edge DP algorithm  which is $(\alpha(D'),\beta(1+\log N) + \beta_s)$-accurate for  $f$ on all dynamic graph sequences with $N$ nodes, maximum degree $D$, and length $T$, where $D'=O(D+\frac{\log T}\eps \log\frac{TN}{\beta_s})$ for all $D\in\N$. The value $D$ does not need to be given to the algorithm.
\end{lemma}

\begin{proof}
     We run the event-level \eedgedp algorithm for estimating the degree sequence from \Cref{lem:estimating_deg} with failure probability set to $\beta_s$ and keep track of the maximum (noisy) degree $\tilde{d}_{\max}^{(t)}$ seen up until time $t$. By \Cref{lem:estimating_deg}, this algorithm is $(\gamma,\beta_s)$-accurate with $\gamma=O(\frac{\log T}\eps \log\frac{TN}{\beta_s})$. Let $\hat{N}=2^{\lceil \log N \rceil}<2N$. Note $\log\hat{N}=\lceil \log N \rceil< \log N +1$. 
     We set $D_j=\gamma+ 2^j$ and $\tau_j=2^j$ for all $j\in[\log \hat{N}]$. For all time steps $t$ such that  $\tau_{j-1}\leq\tilde{d}_{\max}^{(t)}<\tau_j$, we run $\Alg_{D_j}$.

     \paragraph{Accuracy.}  
      Let $\graphseq=(G_1,\dots,G_T)$ and  $\hat{D}^{(t)}$ be the maximum degree of \sr{$(G_1,\dots,G_t)$ for all $t\in[T]$.}
      With probability at least $1-\beta_s$, at all time steps $t\in[T]$, we have that $\hat{D}^{(t)}\in[\tilde{d}_{\max}^{(t)}-\gamma,\tilde{d}_{\max}^{(t)}+\gamma]$. Call that event $C$. Conditioning on $C$, then for $j\in [\log \hat{N}]$,  at all time steps $t$ where we run $\Alg_{D_j}$, we have $\tilde{d}_{\max}^{(t)}\in [\tau_{j-1}, \tau_j)$, and therefore 
      $\hat{D}^{(t)}\in[\tau_{j-1}-\gamma,\tau_j+\gamma)=\left[{D_j}/{2}-\gamma,D_j\right)$.  Conditioning on $C$, with probability at least $1-\beta$, for a fixed $j\in [\log \hat{N}]$, algorithm $\Alg(D_j)$ has error at most $\alpha(D_j)$ for all time steps $t$ where we run $\Alg(D_j)$. Thus, by a union bound, for all $j\in [\log \hat{N}]$, we have that with probability at least $1-\beta_s-\beta(1+\log N)$, the error at every time step $t$ where we run  $\Alg(D_j)$ is bounded by $\alpha(D_j)=\alpha(O(\hat{D}^{(t)}+\gamma))=\alpha(O(\hat{D}^{(T)}+\gamma))$.
     
     \paragraph{Privacy.} The algorithm implicitly partitions $[T]$ into intervals $I_1,\dots, I_{\log \hat{N}}$, such that we run $\Alg_{D_j}$ on $I_j$ for $j\in[\log \hat{N}]$. Call $\Alg_{\mathrm{Int}}$ the algorithm which takes as input a graph sequence $\graphseq$ and outputs the interval partition. Since it is post-processing of the algorithm from \Cref{lem:estimating_deg}, $\Alg_{\mathrm{Int}}$ is event-level \eedgedp.
     
     Let $\graphseq=(G_1,\dots,G_T)$ and $\graphseq'=(G'_1,\dots,G'_T)$ be two neighboring input sequences. 
     Let $\mathcal{I}=(I_1,\dots, I_{\log \hat{N}})$ be an interval partition such that for all $j\in [\log \hat{N}]$, the interval $I_j=[t_{j},t_{j+1})$ and the maximum degrees of $G_t$ and $G'_t$ 
     are bounded by $D_j$ for all $t\in I_j$. 
     We get $$\Alg_{D_j}(G_{t_j}, \dots, G_{t_{j+1}-1})\approx_{\eps,\delta}\Alg_{D_j}(G'_{t_j}, \dots, G'_{t_{j+1}-1})$$ for all $j\in [\log \hat{N}]$, by definition of $\Alg_{D_j}$. Thus, $$\left(\Alg_{D_j}(G_{t_j}, \dots, G_{t_{j+1}-1})\right)_{j\in[\log \hat{N}]}\approx_{\eps',\delta'}\left(\Alg_{D_j}(G'_{t_j}, \dots, G'_{t_{j+1}-1})\right)_{j\in[\log \hat{N}]},$$ where $\eps'=\eps(1+\log N)$ and $ \delta'=\delta(1+\log N)$ by \Cref{lem:composition_theorem}.
     
     Let $\Alg_{\mathrm{all }D}$ be the algorithm that takes as input a sequence $\graphseq$ and an interval partition $\mathcal{I}=(I_1,\dots, I_{\log \hat{N}})$, and runs $\Alg_{D_j}$ on $G_{t_j},\dots, G_{t_{j+1}-1}$ for each $I_j=[t_j,t_{j+1})$ and $j\in [\log \hat{N}]$. Our full algorithm can be seen as $\Alg(\graphseq):= \Alg_{\mathrm{all }D}(\graphseq, \Alg_{\mathrm{Int}}(\graphseq))$. Let $B$ be the set of all interval sequences $\mathcal{I}=(I_1,I_2,\dots, I_{\log \hat{N}})$, where $I_j=[t_{j},t_{j+1})$ for all $j\in[\log \hat{N}]$, such that the maximum degrees of $G_{t_j}, \dots, G_{t_{j+1}-1}$ and $G'_{t_j}, \dots, G'_{t_{j+1}-1}$ are bounded by $D_j$, for all $j\in [\log \hat{N}]$. Let $H$ be the set of intervals where the degree bound is violated for $\graphseq$, and $H'$ be the set of intervals where the degree bound is violated for $\graphseq'$, such that $B= \overline{H\cup H'}$. By the properties of $\Alg_{\mathrm{Int}}$, we have $\Pr[\Alg_{\mathrm{Int}}(\graphseq)\in H]\leq \beta_s$ and $\Pr[\Alg_{\mathrm{Int}}(\graphseq)\in H']\leq e^{\eps}\Pr[\Alg_{\mathrm{Int}}(\graphseq')\in H']\leq e^{\eps}\beta_s$. Thus, by a union bound, $\Pr[\Alg_{\mathrm{Int}}(\graphseq)\in H\cup H']\leq \beta_s+e^{\eps}\beta_s$ and therefore $\Pr[\Alg_{\mathrm{Int}}(\graphseq)\in B]\geq 1-\beta_s-e^{\eps}\beta_s$. For every $\out\in\range(\Alg)$, 
     \begin{align*}
     \Pr[\Alg(\graphseq)\in \out]&=\Pr[\Alg_{\mathrm{all }D}(\graphseq, \Alg_{\mathrm{Int}}(\graphseq))\in\out]
     \\&\leq \Pr[\Alg_{\mathrm{all }D}(\graphseq, \Alg_{\mathrm{Int}}(\graphseq))\in\out | \Alg_{\mathrm{Int}}(\graphseq)\in B]\cdot\Pr[\Alg_{\mathrm{Int}}(\graphseq)\in B]+\beta_s(1+e^{\eps}).
     \end{align*}

     We further bound the first summand:
     \begin{align*}
     &\Pr[\Alg_{\mathrm{all }D}(\graphseq, \Alg_{\mathrm{Int}}(\graphseq))\in\out | \Alg_{\mathrm{Int}}(\graphseq)\in B]\cdot\Pr[\Alg_{\mathrm{Int}}(\graphseq)\in B]
     \\ &=\sum_{\mathcal{I}\in B} \Pr[\Alg_{\mathrm{all }D}(\graphseq, \mathcal{I})\in\out] \Pr[\Alg_{\mathrm{Int}}(\graphseq)=\mathcal{I}]
     \\ &\leq \sum_{\mathcal{I}\in B}\left(e^{\eps(\log {N}+1)}\Pr[\Alg_{\mathrm{all }D}(\graphseq', \mathcal{I})\in\out]+\delta(\log {N}+1)\right)\Pr[\Alg_{\mathrm{Int}}(\graphseq)=\mathcal{I}]
     \\ &\leq \sum_{\mathcal{I}\in B}e^{\eps(\log {N}+2)}\Pr[\Alg_{\mathrm{all }D}(\graphseq', \mathcal{I})\in\out]\Pr[\Alg_{\mathrm{Int}}(\graphseq')=\mathcal{I}]+\delta(\log {N}+1) 
     \\&=e^{\eps(\log N+2)}\Pr[\Alg_{\mathrm{all }D}(\graphseq', \Alg_{\mathrm{Int}}(\graphseq'))\in\out | \Alg_{\mathrm{Int}}(\graphseq')\in B]\Pr[\Alg_{\mathrm{Int}}(\graphseq')\in B]+\delta(\log N+1)
     \\&\leq e^{\eps(\log N+2)}\Pr[\Alg(\graphseq')\in \out]+\delta(\log N+1),
     \end{align*}
    The first inequality follows since for each $\mathcal{I}\in B$, we have that $\Alg_{\mathrm{all }D}(\graphseq, \mathcal{I})\approx_{\eps',\delta'}\Alg_{\mathrm{all }D}(\graphseq', \mathcal{I})$ with $\eps'=\eps(\log N+1)$ and $\delta'=\delta(\log N+1))$, as argued above. The second inequality follows since $\Alg_{\mathrm{Int}}$ is $\eps$-DP.

    Together, this gives
    \begin{align*}
         \Pr[\Alg(\graphseq)\in \out]\leq e^{\eps(\log N+2)}\Pr[\Alg(\graphseq')\in \out]+\delta(\log N+1)]+\beta_s(1+e^{\eps}). 
    \end{align*}
\end{proof}

\begin{proof}[Proof of Theorem~\ref{thm:d-restricted}]
    Apply \Cref{lem:d-rest} with $\eps=\frac{\eps'}{\log N +2}$, $\delta=\frac{\delta'}{2(1+\log N)}$, and $\beta=\beta_s=\min\left(\frac{\delta'}{2+2e^{\eps'}},\frac{\beta'}{2+\log N}\right).$ Note that $\frac{\delta'}{2+2e^{\eps'}}\leq \frac{\delta'}{2+2e^{\eps}}$.
\end{proof}

\subsection{Algorithms Based on Recomputing at Regular Intervals}\label{sec:recomputing-strategy}

In this section, we collect error bounds which follow from the \emph{recomputing strategy}, i.e., recomputing the answer every fixed number of time steps and returning the most recently computed answer at every time step.
First, we prove Theorem~\ref{thm:ub-low_sensitivity} on item-level algorithms for functions with known sensitivity. It generalizes Theorems E.2 and E.5 in \cite{JainRSS23} (stated there for functions of sensitivity 1) to higher-dimensional functions and adapts them for graph functions.

\begin{theorem}[Releasing low sensitivity functions]\label{thm:ub-low_sensitivity}
    Let $V$ be a set of nodes, $r>0$ be a range parameter, and $k\in\N$ be the dimension. Let  $f$ be a function $f:\allgraphs{V}\rightarrow$ $[0,r]^k$ of $L_p$-sensitivity $\Delta_p$ (w.r.t.\ the edge-neighboring metric in static graphs) for all $p\in\{1,2\}$. For all $\eps>0$, $\delta\in [0,1)$, $\beta\in(0,1)$, there exists an item-level \ededgedp algorithm which is $(\alpha,\beta)$-accurate for estimating
    $f$ on all dynamic graph sequences of length $T\in \N$, satisfying the following accuracy guarantee: 
    \begin{enumerate}
        \item 
        if $\delta=0$, then $\alpha=O\left(\min\left(\Delta_1\cdot\sqrt{\frac T{\eps}\ln\frac {Tk}\beta},r\right)\right)$;
        
        \item if $\delta>0$, then $\alpha=O\left(\min\left(\sr{\Delta_2}
        \cdot\sqrt[3]{\frac T{\eps^2}(\ln\frac {Tk}\beta)\ln\frac 1\delta},r\right)\right)$.
    \end{enumerate}
    \end{theorem}

    \begin{proof}
    The second term in the minimum expression (for both items) comes from the trivial algorithm that always outputs $f(G_0)$, where $G_0=(V,\emptyset),$ without looking at the data. It remains to analyze the first term.
    For all $p\in\{1,2\}$ and every function $f$ with $L_p$-sensitivity $\Delta_p$, the rescaled function $f'=f/\Delta_p$ has sensitivity 1. Therefore, it suffices to 
    analyze the first term in the minimum
    for functions with $\Delta_p=1$.
    
    The algorithm for the first term uses an edge-DP algorithm to recompute the function value every $B$ time steps, and does not update the output in between. That is, we divide the time line into \emph{blocks} of the form $[1,B],[B+1,2B],\dots,[(\lceil \frac{T}{B}\rceil -1)B+1,T]$. Denote $t_1,\dots, t_{\lceil T/B\rceil}$ the last time step in each block, i.e., $t_i=i\cdot B$ for all $i\in[\lceil T/B\rceil-1]$ and $t_{\lceil T/B\rceil}=T$. Since $f$ has $L_p$-sensitivity 1, and all $G_i$, $G_{i+1}$ are edge-neighboring, we have that $\|f(G_{t_{i+1}})-f(G_{t_{i}})\|_{\infty}\leq B$ for all $i\in[\lceil T/B\rceil-2]$. We first set the output to $0$. At every time step $t_i$, we compute an estimate of $f(G_{t_i})$ using a differentially private mechanism, and within a block, we do not update the output. That is, for $\eps$-differential privacy, we run the Laplace mechanism on $\lceil\frac T B\rceil$ inputs $G_{t_1},\dots,G_{t_{\lceil T/B\rceil}}$, and for every $t_i<t<t_{i+1}$, we output the same value as in the previous time step. Since for each of these inputs and every item-level edge-neighboring sequence $G'_1,\dots, G'_T$, we have that $G_{t_j}$ and $G'_{t_j}$ are neighboring, the $L_1$-sensitivity of $(f(G_{t_1}),\dots, f(G_{t_{\lceil T/B\rceil}}))$ is $\lceil\frac T B\rceil$. The value of $\alpha$ follows by balancing $B$ with the error of the Laplace mechanism: By \Cref{lem:Laplacemech}, the Laplace mechanism has error at most $\lceil\frac T B\rceil\frac 1 \eps \ln\frac{kT}{\beta}$. Setting $B=\Theta\left(\sqrt{\frac{T}{\eps}\ln\frac{Tk}\beta}\right)$ gives the claimed bound $\alpha$. Similarly, for $(\eps,\delta)$-differential privacy, we run the Gaussian mechanism on $\lceil \frac T B\rceil$ inputs $G_{t_1},\dots,G_{t_{\lceil T/B\rceil }}$. Since for each of these inputs and every item-level neighboring sequence $G'_1,\dots, G'_T$, we have that $G_{t_j}$ and $G'_{t_j}$ are neighboring, the $L_2$-sensitivity of $(f(G_{t_1}),\dots, f(G_{t_{\lceil T/B\rceil}}))$ is $\sqrt{\lceil T/B\rceil}$. The value of $\alpha$ follows by balancing $B$ with the error of the Gaussian mechanism: By \Cref{lem:gaussianmech}, the Gaussian mechanism has error at most $\sqrt{\lceil T/B\rceil}\frac{2}\eps \sqrt{\ln(2/\delta)\ln(2Tk/\beta)}$. Setting $B=\Theta(\sqrt[3]{\frac T{\eps^2}(\ln\frac {Tk}\beta)\ln\frac 1\delta})$ gives the claimed bound $\alpha$.
    \end{proof}

Theorem~\ref{thm:ub-low_sensitivity} immediately yields the following corollary for functions with constant sensitivity, including \edgecount, \textnormal{HighDegree}, \maxmatch, \mincut ($\mincutmath$), \conncomp, \degreehist,\degreelist.

\begin{corollary}[Item-level algorithms]\label{cor:ub-low-sensitivity-item-level} 
Let $\cC$ be the class of functions of constant sensitivity from graphs to $[0,N]$ (that includes $\highdegreemath{\tau}\forall\tau\in\N,\maxmatchmath,\mincutmath,$ and $\conncompmath$) and $\cC_N$ be the class of functions from graphs to real-valued vectors of length $O(N)$ with entries in $[0,N]$ (that includes \degreehist and \degreelist).
For all $\eps>0,\delta\in[0,1),\beta\in(0,1)$, $\tau\in\N$ and $T\in \N$ and $f\in\cC\cup\cC_N$, there exists an item-level \ededgedp algorithm which is $(\alpha,\beta)$-accurate
    for $f$ on all 
    dynamic graph sequences of length $T$ such that 
    \sr{$\alpha=O(\min(\alpha',N^2))$ for $\edgecountmath$ and $\alpha=O(\min(\alpha',N))$ for all $f\in \cC\cup\cC_N$, where
     \begin{enumerate}
        \item 
        if $\delta=0$, then $\alpha'=\sqrt{\frac{T}{\eps}\ln\frac{T}{\beta}}$ for $\edgecountmath$ and all $f\in\cC$, and $\alpha'=\sqrt{\frac{T}{\eps}\ln\frac{TN}{\beta}}$ for all $f\in\cC_N$.
        \item if $\delta>0$, then $\alpha'=\sqrt[3]{\frac T{\eps^2}(\ln\frac {T}\beta)\ln\frac 1\delta}$ for $\edgecountmath$ and all $f\in\cC$, and $\alpha'=\sqrt[3]{\frac T{\eps^2}(\ln\frac {TN}\beta)\ln\frac 1\delta}$ for all $f\in\cC_N$.
    \end{enumerate}
    }

    \end{corollary}

By Theorem~\ref{thm:ub-low_sensitivity}, there exists a $D$-restricted item-level 
(and therefore also event-level), $\eps$-edge-DP algorithm for fully dynamic triangle count which is $(\alpha,\beta)$-accurate with $\alpha=O\left(D\sqrt{\frac{T}{\eps}\ln\frac{T}{\beta}}\right)$, and a $D$-restricted, event-level, $(\eps,\delta)$-edge-DP algorithm for fully dynamic triangle count which is $(\alpha,\beta)$-accurate with $\alpha=O\left(D\cdot\sqrt[3]{\frac T{\eps^2}(\ln\frac {T}\beta)\ln\frac 1\delta}\,\right)$. (Observe that when $D$ is set to $N$, every $D$-restricted algorithm becomes edge-DP because the promise on the degree becomes vacuous.)
Together with \Cref{thm:d-restricted}, we get the following corollary. 

\begin{corollary}[Triangle count, recomputing bound]\label{cor:upperbound_triangle_item-level}
    Let $\eps>0$ and $\delta,\beta\in(0,1)$. Let $T,N,D\in \N$. There exists an 
    event-level \ededgedp algorithm which is $(\alpha,\beta)$-accurate for \trianglecount on all dynamic graph sequences with $N$ nodes, maximum degree $D$ and length $T$, where $$\alpha=O\left(T^{1/3}D\cdot \mathrm{poly}\left(\frac 1 \eps\log\frac{TN}{\delta\beta}\right)\right).$$
There exists an
    item-level \ededgedp algorithm which is $(\alpha,\beta)$-accurate for \trianglecount on all dynamic graph sequences with $N$ nodes and length $T$, where 
    \begin{enumerate}
 \item if $\delta=0$, then  $\alpha=O\left(\min\left(\min(T,N)\cdot\sqrt{\frac{T}{\eps}\ln\frac{T}{\beta}},N^3\right)\right).$ 
 \item if $\delta>0$, then $\alpha=O\left(\min\left(\min(T,N)\cdot\sqrt[3]{\frac T{\eps^2}(\ln\frac {T}\beta)\ln\frac 1\delta},N^3\right)\right)$.
 \end{enumerate}
    
    \end{corollary}

\section{A Lower Bound Framework for DP Fully Dynamic Algorithms}\label{sec:lb-framework}
We start this section by defining terminology and gadgets used in our lower bound framework. In \Cref{sec:event-level-framework} (respectively, \Cref{sec:item-level-framework}), we present a general theorem for proving event-level (respectively, item-level) lower bounds for fully dynamic graph algorithms.

Let $\mathcal{G}$ denote the set of all graphs, and let $f:\mathcal{G}\rightarrow \R$ be a real-valued graph function.

\begin{definition}[Distinguishing gadget] Let $H=(V',E')$ be a graph and $e_1,e_2\in V'\times V'$. Set $n_g=|V'|$ and $m_g=|E'|$.  The pair $(H,e_1)$ is called a  {\em 1-edge distinguishing gadget} for $f$ of size $(n_g,m_g)$ and weight $w$ if either 
\begin{itemize}
    \item  $e_1\notin E'$ and $f((V',E'\cup\{e_1\}))=f(H)+w$ or
    \item $e_1\in E'$ and $f((V',E'\setminus\{e_1\}))=f(H)+w$.
\end{itemize}
The triple $(H,e_1,e_2)$ is called a {\em 2-edge distinguishing gadget} for $f$ of size $(n_g,m_g)$ and weight $w$, if either
\begin{itemize}
    \item $e_1,e_2\notin E'$ and $f((V',E'\cup\{e_1,e_2\}))=f(H)+w$ and $f((V',E'\cup\{e_1\}))=f((V',E'\cup\{e_2\}))=f(H)$ or
    \item $e_1,e_2\in E'$ and $f((V',E'\setminus\{e_1,e_2\}))=f(H)+w$ and $f((V',E'\setminus\{e_1\}))=f((V',E'\setminus\{e_2\}))=f(H)$.
\end{itemize} 
If there exists a distinguishing gadget for $f$, we say that $f$ has a distinguishing gadget.
\end{definition}

\begin{figure}[t]
    \centering
    \includegraphics[width=0.5\textwidth,trim={0 6cm 0 0},clip]{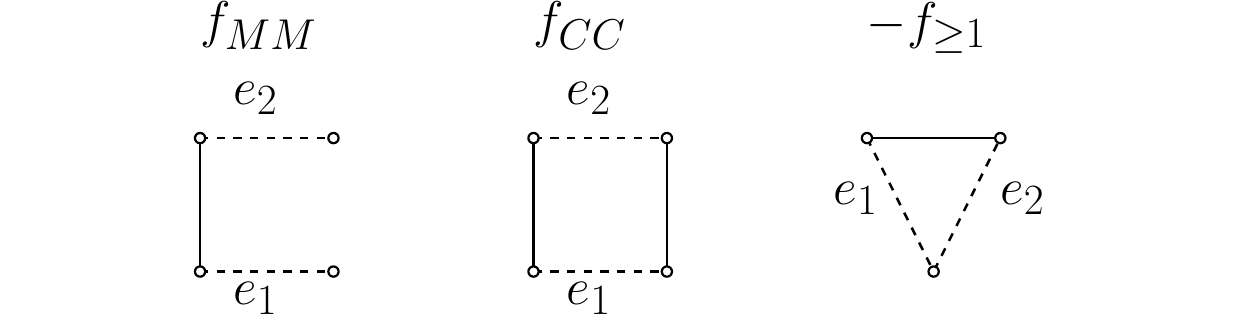}
     \caption{2-edge distinguishing gadgets for $\maxmatchmath,\conncompmath$ and $-\highdegreemath{1}$ of constant size and weight. Note that for $\maxmatchmath$, we have $e_1,e_2\notin E'$, and for $\conncompmath$ and $-\highdegreemath{1}$, we have $e_1,e_2\in E'$.}
    \label{fig:gadgets_gen}
\end{figure}

Examples of distinguishing gadgets are illustrated in Figure~
\ref{fig:gadgets_gen}.

\begin{lemma}\label{lem:gadget-conversion}
    If $f$ has a 2-edge distinguishing gadget of size $(n_g,m_g)$ and weight $w$, then $f$ also has a 1-edge distinguishing gadget of size $(n_g,m_g+1)$ or $(n_g,m_g-1)$ and weight $w$.
\end{lemma}

Our framework applies to additive graph functions, defined next.    
\begin{definition}[Additive graph function]
    A function $f$ is \emph{additive} if for each graph $G$ consisting of the connected components $C_1,\dots, C_m$, we have 
$f(G)=\sum_{i\in[m]}f(C_i)$.
\end{definition}

\subsection{Framework for Event-Level Lower Bounds}\label{sec:event-level-framework}
In this section, we state and prove \Cref{thm:general_event_lower} that encapsulates our lower bound framework for event-level DP. We also apply our framework to specific problems, stating the resulting lower bounds in \Cref{cor:lb-event-level}.

\begin{theorem}[Event-level lower bound for 2-edge distinguishing]\label{thm:general_event_lower}
   Let $f:\allgraphs{V}\rightarrow \R$ be an additive function 
   with
   a 2-edge distinguishing gadget 
   of size $(n_g,m_g)$ and weight $w$ for  some $n_g,m_g,w\in\N$.  Then, for all sufficiently large $T, N\in\N$, every event-level, $(1,1/3)$-edge DP algorithm which is $(\alpha,0.01)$-accurate for $f$  on all dynamic graph sequences with $N$ nodes and length $T$ has error $\alpha=\Omega\left(w\cdot\min\Big(T^{1/4},\sqrt{\frac{T}{m_g}},\sqrt{\frac{N}{n_g}}\Big)\right)$.
\end{theorem}

We prove Theorem~\ref{thm:general_event_lower} via a  reduction from the inner product problem, defined next.
\begin{definition}[Inner product]\label{def:inner-product}
In the inner product problem, the data universe is $\uni=\{0,1\}$. An inner product query on $\uni^n$ is defined by a vector $q\in\{0,1\}^n$ and maps $y\in\uni^n$ to $q\cdot y$. The problem $\innerprod{n}{k}$ is the problem of answering $k$ inner product queries $q^{(1)},\dots, q^{(k)}\in\{0,1\}^n$.
\end{definition}

The following formulation of the lower bound is based on \cite{JainKRSS23}, which is in turn based on \cite{DinurN03,DworkMT07,MirMNW11,De12}.
\begin{lemma}[Inner Product lower bound]\label{lem:inner_prod}
   There exist constants $c_1 \geq 1$ and $c_2 > 0$ such that, for all large enough $n\in\N$, every $(1,1/3)$-DP and $(\alpha,0.01)$-accurate algorithm for $\innerprod{n}{c_1 n}$  satisfies $\alpha>c_2\sqrt{n}$.

\end{lemma}

\begin{lemma}[Reduction from Inner Product for event-level]\label{lem:red:general_event}
    Let $n,k\in\N$. Let $y\in\{0,1\}^n$ and $Q=(q^{(i)})_{i\in[k]}$, where each $q^{(i)}\in\{0,1\}^n$, be a sequence of $k$ inner product queries. Then for all $n_g,m_g,w\in\N$ and each additive function $f:\allgraphs{V}\rightarrow \R$ with a 2-edge distinguishing gadget of size $(n_g,m_g)$ and weight $w$, there exists a transformation from $(y,Q)$ to a dynamic graph sequence $(G_1,\dots, G_T)$ of $N$-node graphs, where $T=(m_g+2k)n$ 
    and $N=n_gn$, such that
    \begin{itemize}
        \item For neighboring $y,y'\in\{0,1\}^n$ and the same query sequence $Q$, the transformations of $(y,Q)$ and $(y',Q)$ give event-level,  edge-neighboring graph sequences;
        \item Let $t_0=(m_g+1)n$. For all ${\ell}\in[k]$ and 
        $t_{\ell}=(m_g+2{\ell})n$,
        we have $f(G_{t_{\ell}})-f(G_{t_0})=wy\cdot q^{({\ell})}$.
    \end{itemize}
\end{lemma}
\begin{proof}
    Let $H=(V',E')$ and $(H,e_1,e_2)$ be a 2-edge distinguishing gadget of size $(n_g,m_g)$ of weight $w$ for $f$. W.l.o.g.\ assume $e_1,e_2\notin E'$. We define a graph sequence on $n\cdot n_g$ nodes as follows:
    In an \emph{initialization phase}, we build $n$ copies of $H$, denoted $H^{(1)}, \dots, H^{(n)}$, using $m_gn$  time steps. For a vertex pair $e$ in $H$ and $j\in[n]$, let $e^{(j)}$ denote the copy of $e$ in $H^{(j)}$.
    For every $j\in[n]$, if $y[j]=1$, insert edge $e_1^{(j)}$ 
    at time step $m_gn+j$ (otherwise, do nothing at that time step). 
    
   For each query vector $q^{({\ell})}$, where  ${\ell}\in[k]$, we first have an \emph{insertion phase}, where 
   for all $j\in[n]$, if $q^{({\ell})}[j]=1$, then we insert edge $e_2^{(j)}$ 
   into $H^{(j)}$ 
    at time $t_0+2({\ell}-1)n+j$; otherwise, we do nothing at that time step. For all but the last query, the insertion phase is followed by the \emph{deletion phase}, where the edges that were inserted in the insertion phase are deleted over $n$ time steps.

    \emph{Correctness.} If $y,y'\in\{0,1\}^n$ are neighboring input vectors differing in coordinate $i^*$, then the resulting graph sequences for $(y,Q)$ and $(y',Q)$ differ only in the absence or presence of the initial insertion of $e_1^{(i^*)}$. Thus, they are event-level,  edge-neighboring. 
    The insertion phase for the ${\ell}$th query ends at time step $t_{\ell}=(m_g+2{\ell})n$.
   At that time step, for all $i\in[n]$, the copy $H^{(i)}$ contains both $e_1^{(i)}$ and $e_2^{(i)}$ if and only if $y[i]=q[i]=1$. Thus, $f(G_{t_{\ell}})=f(G_{t_0})+w\cdot y\cdot q$.

    \emph{Analysis.} The initialization phase uses $(m_g+1)n$ time steps. Each of the $k$ insertion phases and $k-1$ deletion phases uses $n$ time steps. Thus, $T=(m_g+2k)n$. There are $n\cdot n_g$  nodes in the graph.
\end{proof}

\begin{proof}[Proof of Theorem~\ref{thm:general_event_lower}]
Let $c_1,c_2$ be as in \Cref{lem:inner_prod}. Given $T$ and $N$, set $n=\min\left(\sqrt{\frac{T}{4c_1}},\frac{T}{2m_g},\frac{N}{n_g}\right)$. Then for each $y\in\{0,1\}^n$ and $k$-query sequence $Q$, where $k=c_1n$, the transformation in \Cref{lem:red:general_event} gives a sequence with at most $2c_1n^2+m_gn\leq T$
time steps and at most $N$ nodes. We can add dummy nodes and time steps to get a sequence of exactly $T$ time steps and $N$ nodes. If we can estimate $f(G_t)$ 
up to error $\alpha$ at all time steps with probability at least $1-0.01$, then with the same probability, we can estimate $y\cdot q^{(m)}=\frac{f(G_{t_m})-f(G_{t_0})}{w}$ up to error 
$\frac{2\alpha}w$ for all $m\in[k]$. Thus,  $\alpha\geq w\cdot \sqrt{n}\cdot\frac{c_2}{2}=\Omega\left(w\cdot\min\Big(T^{1/4},\sqrt{\frac{T}{m_g}},\sqrt{\frac{N}{n_g}}\Big)\right)$.
\end{proof}

\subsubsection{Event-Level Lower Bounds for Specific Problems}\label{sec:lb-event-level-specific-problems}

 \Cref{thm:general_event_lower} yields the following corollary  for $\maxmatch,$ $\conncomp,$ $\highdegree{\tau},$ and $\degreehist$. 

\begin{corollary}[Event-level lower bounds]\label{cor:lb-event-level}
   Let $\tau\in\N$ and $f\in\{\maxmatchmath,\conncompmath,\highdegreemath{\tau},\degreehistmath\}$.  Then, for all sufficiently large $T, N\in\N$, every event-level, $(1,1/3)$-edge DP algorithm which is $(\alpha,0.01)$-accurate  for $f$  on all dynamic graph sequences with $N$ nodes and length $T$ has error  $\alpha=\Omega\left(\min\Big(T^{1/4},\sqrt{N}\Big)\right)$.

\end{corollary}
\begin{proof}
Using the gadgets illustrated in \Cref{fig:gadgets_gen} in  \Cref{thm:general_event_lower} gives the stated  bound for $\maxmatch,$ $\conncomp,$ and $\highdegree{1}.$ The lower bound for $\degreehist$ is inherited from the lower bound on $\highdegree{1}$: 
one of the entries in the degree histogram of a graph $G=(V,E)$ is the number of isolated nodes, which is $|V|-\highdegreemath{1}(G)$.
%

Finally, the lower bound for $\highdegreemath{\tau}$ for $\tau\geq 2$ is inherited from the lower bound for $\highdegreemath{1}$. Fix $\tau\geq 2$. We can reduce from fully dynamic $\highdegreemath{1}$ to fully dynamic $\highdegreemath{\tau}$ as follows. Let $\graphseq=(G_1,\dots,G_T)$ be a dynamic graph sequence on vertex set $V$ of size $N\geq 2$. We construct a new graph sequence $\graphseq_\tau$ on nodes $V\cup U$, where $U$ is disjoint from $V$ and $|U|=\tau-1$. We initialize $\graphseq_\tau$  by inserting a complete bipartite graph on $U\times V$ and a complete graph on $U$. This takes $t_0=(\tau-1)(N+\frac{\tau-2}{2})$ time steps. Then we proceed with the same updates on $V$ as in $\graphseq$. The resulting sequence $\graphseq_\tau$ is on graphs with $N_\tau=N+\tau$ nodes and has length $T_\tau=t_0+T.$ 
Let $\graphseq_\tau=(H_1,\dots,H_{T_\tau})$.

In $H_{t_0}$ (at the end of initialization), all vertices from $U$ have degree $\tau+|V|-2\geq \tau$, and all vertices from $V$ have degree $\tau-1$. For all $t\in[T]$, we have $\highdegreemath{1}(G_t)=\highdegreemath{\tau}(H_{t_0+T})-(\tau-1)$. 
This transformation preserves event-level (as well as item-level) edge-neighboring relationship. Since $N_\tau=\Theta(N)$ and $T_\tau=\Theta(T)$, the additive error lower bound for $\highdegreemath{1}$ applies to $\highdegreemath{\tau}$.
\end{proof}

\subsection{Framework for Item-Level Lower Bounds}\label{sec:item-level-framework}
In this section, we state and prove \Cref{thm:general_item_lower}  that encapsulates our lower bound framework for item-level DP. 
We also apply our framework to specific problems in \Cref{sec:lb-item-level-specific-problems}.

\begin{theorem}[Item-level lower bound for 1-edge distinguishing]\label{thm:general_item_lower}
   Let  $f:\allgraphs{V}\rightarrow \R$ be an additive function with a 1-edge distinguishing gadget of size $(n_g,m_g)$ and weight $w$ for some $n_g,m_g,w\in\N$.  Then, for all $\eps\in(0,1]$, $\delta\in[0,1)$, and sufficiently large $T, N\in\N$, every item-level, $(\eps,\delta)$-edge DP algorithm which is $(\alpha,\frac{1}{3})$-accurate for $f$  on all dynamic graph sequences with $N$ nodes and length $T$ satisfies 
   \begin{enumerate}
       \item if $\delta>0$ and $\delta=o(\frac 1 T)$, then $\alpha=\Omega\left(w\cdot\min\left(\frac{T^{1/3}}{\eps^{2/3}\log^{2/3} T},\frac{T}{m_g}, \frac{N}{n_g}\right)\right)$.
       \item if $\delta=0$, then $\alpha=\Omega\left(w\cdot\min\left(\sqrt{\frac{T}{\eps}},\frac{T}{m_g},\frac{N}{n_g}\right)\right)$.
   \end{enumerate}
\end{theorem}

We prove Theorem~\ref{thm:general_item_lower} by reducing from the Marginals problem (Definition~\ref{def:marginals}). 
\begin{lemma}[Reduction from Marginals for item-level]\label{lem:red:general_item}
    Let $n,d\in\N$ and $Y\in\{\{0,1\}^d\}^n$. Then for all $n_g,m_g,w\in\N$ and each additive function $f:\allgraphs{V}\rightarrow \R$ with a 1-edge distinguishing gadget of size $(n_g,m_g)$ and weight $w$, there exists a transformation from $Y$ 
    to a dynamic graph sequence $(G_1,\dots, G_T)$ of $N$-node graphs, where $T=(m_g+2d)n$ and $N=n_gn$, such that
    \begin{itemize}
        \item The transformations of neighboring $Y,Y'\in\{\{0,1\}^d\}^n$ give item-level,  edge-neighboring graph sequences;
        \item Let $t_0=(m_g+1)n$. For all $j\in[d]$ and 
        $t_j=(m_g+2j)n,$
        we have $f(G_{t_j})-f(G_{t_0})=w\sum_{i=1}^nY_i[j]$.
    \end{itemize}
\end{lemma}
\begin{proof}
    Let $H=(V',E')$ and $(H,e_1)$ be a 1-edge distinguishing gadget of size $(n_g,m_g)$ of weight $w$ for $f$. W.l.o.g.\ assume $e_1\notin E'$. We define a graph sequence on $n\cdot n_g$ nodes as follows:
    In an \emph{initialization phase}, we build $n$ copies of $H$, denoted $H^{(1)}, \dots, H^{(n)}$, using $m_gn$  time steps. 
    For a vertex pair $e$ in $H$ and $i\in[n]$, let $e^{(i)}$ denote the copy of $e$ in $H^{(i)}$.
    
    For all $j\in[d]$, we first have an \emph{insertion phase}, where for all $i\in[n]$, if $Y_i[j]=1$, then we insert $e_1^{(i)}$ into $H^{(i)}$ at time $m_gn+2(j-1)n+i$; otherwise, we do nothing at that time step. For all but the last phase, the insertion phase is followed by a \emph{deletion phase}, where the edges that were inserted in the insertion phase are deleted.

    \emph{Correctness.} If $Y$ and $Y'$ are neighboring input 
datasets
 differing in the row $i^*\in[n]$, then the resulting graph sequences 
    differ only in insertions and deletions related to $e_1^{(i^*)}$. Thus, they are item-level,  edge-neighboring. The insertion phase for the $j$th query ends at time step $t_j=(m_g+1)n+2(j-1)n+n$. At that time step, the $H^{(i)}$ contains $e_1^{(i)}$  if and only if $Y_i[j]=1$. Thus, $f(G_{t_j})=f(G_{t_0})+w\cdot \sum_{i\in [n]}Y_i[j]$.

    \emph{Analysis.} The initialization phase uses $(m_g+1)n$ time steps. Each of the $d$ insertion phases and $d-1$ deletion phases uses $n$ time steps. Thus, $T=(m_g+2d)n$. There are $N=n\cdot n_g$  nodes in the graph.
\end{proof}

   \begin{proof}[Proof of Theorem~\ref{thm:general_item_lower}] 
   {\bf Case $\delta>0$:} Given $T$ and $N$, set $d=\big\lfloor\left(T\eps\log{T\eps}\right)^{2/3}\big\rfloor$ and $n=\lfloor\min\left(\frac{\sqrt{d}}{6\eps\log d},\frac{T}{2m_g}, \frac{N}{n_g}\right)\rfloor$. The reduction from \Cref{lem:red:general_item} 
   gives a sequence with at most $N$ nodes and $2nd+T/2$ updates, where 
   $$2nd\sr{\leq}\frac{d^{3/2}}{3\eps\log d}\leq\frac{{T\eps}\log ({T\eps})}{3\eps\cdot\frac{2}{3}\log({T\eps}\log ({T\eps}))}\leq \frac T2.$$
   We can add  dummy nodes and dummy time steps to get a dynamic $N$-node graph sequence of length $T$. If we can estimate $f(G_t)$ with error up to $\alpha$ at all time steps with probability at least $2/3$, then for all $j\in[d]$, we can estimate the $j$th marginal given by $\frac{1}{n}\sum_{i\in n}Y_i[j]=\frac{f(G_{t_j})-f(G_{t_0})}{nw}$ up to  error $\frac{2\alpha}{nw}$. \Cref{lem:marginals}
    gives $\alpha=\Omega\left(wn\cdot\min\left(\frac{\sqrt{d}}{n\eps\log d},1\right)\right)=\Omega\left(w\cdot\min\left(\frac{T^{1/3}}{\eps^{2/3}\log^{2/3} T},\frac{T}{m_g}, \frac{N}{n_g}\right)\right)$.

    {\bf Case $\delta=0$:} Given  $T$ and $N$, set $d=\lfloor\sqrt{T\eps}\rfloor$ and $n=\lfloor\min\left(\frac{T}{(m_g+1)},\frac{N}{n_g}, \frac{d}{4\eps}\right)\rfloor$. The reduction from \Cref{lem:red:general_item} gives a sequence with at most $N$ nodes and at most $T$ updates, and we can add dummy nodes and dummy time steps to get a dynamic $N$-node graph sequence of length $T$. By \Cref{lem:marginals}, we have 
    $\alpha=\Omega\left(wn\cdot\min\left(\frac{d}{n\eps},1\right)\right)=\Omega\left(w\cdot\min\left(\sqrt{\frac{T}{\eps}},\frac{T}{m_g},\frac{N}{n_g}\right)\right)$.\end{proof}

\subsubsection{Item-Level Lower Bounds for Specific Problems}\label{sec:lb-item-level-specific-problems}

 Using 
the 2-edge distinguishing gadgets from \Cref{fig:gadgets_gen} together with \Cref{lem:gadget-conversion} on converting 2-edge distinguishing gadgets to 1-edge distinguishing, we immediately get the following corollary for  
$\maxmatch$, $\conncomp$, and $\highdegree{1}$  from \Cref{thm:general_item_lower}. By the same reasoning as in the proof of Corollary~\ref{cor:lb-event-level}, this lower bound also applies to \degreehist and to $\highdegree{\tau}$ for $\tau\geq 2$.

\begin{corollary}[Item-level lower bounds]\label{cor:lb-item-level}
Let $\tau\in\N$ and $f\in\{\maxmatchmath,\conncompmath,\highdegreemath{\tau},\degreehistmath\}$.
 Then, for all $\eps\in(0,1]$, $\delta\in[0,1)$, and sufficiently large $T, N,D\in\N$, every item-level, $(\eps,\delta)$-edge DP algorithm which is $(\alpha,\frac{1}{3})$-accurate  for $f$ on all dynamic graph sequences with $N$ nodes and length $T$ satisfies 
   \begin{enumerate}
       \item if $\delta>0$ and $\delta=o(\frac 1 T)$, then 
       $\alpha=\Omega\left(\min\left(\frac{T^{1/3}}{\eps^{2/3}\log^{2/3} T},T, N\right)\right)$.
       \item if $\delta=0$, then 
       $\alpha=\Omega\left(\min\left(\sqrt{\frac{T}{\eps}},T,N\right)\right)$.
   \end{enumerate}
\end{corollary}

Next, we prove the item-level lower bound for \edgecount.
It cannot be obtained as a direct corollary of \Cref{thm:general_item_lower} because the gadgets we use for this problem overlap, but otherwise it follows a similar proof strategy and yields a similar-looking (albeit not exactly the same) bound.
\begin{theorem}[Item-level lower bound for edge count]\label{thm:lb-edge_item-level}
   For all $\eps\in(0,1]$, $\delta\in[0,1)$, and sufficiently large $T, N\in\N$, every item-level, $(\eps,\delta)$-edge DP algorithm for $\edgecountmath$ which is $(\alpha,\frac{1}{3})$-accurate on all dynamic graph sequences with $N$ nodes and length $T$ satisfies 
   \begin{enumerate}
       \item if $\delta>0$ and $\delta=o(\frac 1 T)$, then $\alpha=\Omega\left(\min\left(\frac{T^{1/3}}{\eps^{2/3}\log^{2/3} T},N^2\right)\right)$.
       \item if $\delta=0$, then $\alpha=\Omega\left(\min\left(\sqrt{\frac{T}{\eps}},N^2\right)\right)$.
   \end{enumerate}
\end{theorem}

As in the general framework of \Cref{thm:general_item_lower}, we reduce from Marginals (recall \Cref{def:marginals}).

\begin{lemma}[Reduction from Marginals to \edgecount]\label{lem:red:edge-count}
    Let $n,d\in\N$ and $Y\in\{\{0,1\}^d\}^n$. There exists a transformation from $Y$ 
    to a dynamic graph sequence $(G_1,\dots, G_T)$ of $N$-node graphs, where $T=2n d$ and $N=\lceil\sqrt{2n+1}\rceil$, such that
    \begin{itemize}
        \item The transformations of neighboring $Y,Y'\in\{\{0,1\}^d\}^n$ give item-level,  edge-neighboring graph sequences;
        \item  For all $j\in[d],$ 
        we have $\edgecountmath(G_{t_j})=\sum_{i=1}^nY_i[j]$, where $t_j=(2j-1)n$.
    \end{itemize}
\end{lemma}
\begin{proof}
    We define a set $V$ of $\lceil\sqrt{2n+1}\rceil$ nodes. 
    $V$ has at least $n$ node pairs. We give them an arbitrary order $e_1,\dots, e_{\lceil\sqrt{2n+1}\rceil^2}$. In particular, there exist the node pairs $e_1,\dots, e_n$.

    For all $j\in[d]$ and all $i\in[n]$, if $Y_i[j]=1$, then we insert $e_i$ at time $2(j-1)n+i$; otherwise, we do nothing at that time step. For all $j\in[d-1]$ and all $i \in[n]$, we delete $e_i$ at time $(2j-1)n+i$. 

    \emph{Correctness.} If $Y$ and $Y'$ are neighboring input 
datasets
differing in row $i^*$, then the resulting graph sequences for $Y$ and $Y'$ differ only in insertions and deletions related to $e_i$. Thus, they are item-level, edge-neighboring. At time step $t_j=(2j-1)n$, the graph $G_{t_j}$ includes an edge $e_i$ if and only if $Y_i[j]=1$. Thus, the total edge count is equal to $\sum_{i=1}^nY_i[j]$. 
    \end{proof}

       \begin{proof}[Proof of \Cref{thm:lb-edge_item-level}] 
   {\bf Case $\delta>0$:} Given $T$ and $N$, set $d=\big\lfloor\left(T\eps\log{T\eps}\right)^{2/3}\big\rfloor$ and $n=\lfloor\min\left(\frac{\sqrt{d}}{3\eps\log d}, N^2\right)\rfloor$. The reduction from \Cref{lem:red:edge-count} 
   gives a sequence with at most $N$ nodes and $2nd$ updates, where 
   $$2nd=\frac{2d^{3/2}}{3\eps\log d}\leq\frac{2{T\eps}\log ({T\eps})}{3\eps\cdot\frac{2}{3}\log({T\eps}\log ({T\eps}))}\leq T.$$
   We can add  dummy nodes and dummy time steps to get a dynamic $N$-node graph sequence of length $T$. If we can estimate $\edgecountmath(G_t)$ with error up to $\alpha$ at all time steps with probability at least $2/3$, then for all $j\in[d]$, we can estimate the $j$th marginal given by $\frac{1}{n}\sum_{i\in n}Y_i[j]=\frac{1}{n}\edgecountmath(G_{t_j})$ up to  error $\frac{\alpha}{n}$. \Cref{lem:marginals}
    gives $\alpha=\Omega\left(n\cdot\min\left(\frac{\sqrt{d}}{n\eps\log d},1\right)\right)=\Omega\left(\min\left(\frac{T^{1/3}}{\eps^{2/3}\log^{2/3} T},N^2\right)\right)$.

    {\bf Case $\delta=0$:} Given  $T$ and $N$, set $d=\lfloor\sqrt{T\eps}\rfloor$ and $n=\lfloor\min\left(N^2, \frac{d}{2\eps}\right)\rfloor$. The reduction from \Cref{lem:red:edge-count} gives a sequence with at most $N$ nodes and at most $T$ updates, and we can add dummy nodes and dummy time steps to get a dynamic $N$-node graph sequence of length $T$. By \Cref{lem:marginals}, we have 
    $\alpha=\Omega\left(n\cdot\min\left(\frac{d}{n\eps},1\right)\right)=\Omega\left(\min\left(\sqrt{\frac{T}{\eps}},N^2\right)\right)$.
\end{proof}

\begin{remark}\label{remark:deg-list}
\degreelist inherits lower bounds from \edgecount, except that $N$ gets replaced with $\sqrt N$. This holds both for event-level and item-level edge DP. This is because there is a direct reduction from fully dynamic $\edgecountmath$ to fully dynamic $\degreelist$ that converts each dynamic graph sequences $\graphseq$ on $N$ nodes of length $T$ to a dynamic graph sequences $\graphseq'$ on $\binom N 2+1$ nodes of the same length. Given a vertex set $V$ of $\graphseq$, the vertex set $V'$ of $\graphseq'$ is defined as $\{u\}\cup \{v_e: e\in V\times V\}$. Then each update on an edge $e$ in $\graphseq$ is replaced with the same type of update on the edge $(u,v_e)$ in $\graphseq'$. At every time step, the degree of $u$ in $\graphseq'$ is equal to the edge count in $\graphseq$. Moreover, this reduction preserves event-level and item-level neighbor relationships.
\end{remark}

\subsection{Framework for Event-Level Output-Determined Lower Bounds}
In this section, we show better 
lower bounds for a special class of algorithms, called {\em output-determined}, namely, those that depend only on the output sequence of the true function: If two sequences have the same output sequences, then the algorithm behaves in (roughly) the same way.
Note that all algorithms \sr{for $\degreehistmath,\highdegreemath{\tau},\maxmatchmath, \conncompmath$} proposed in this work are in that class. The lower bounds we present in this section show that for $\degreehistmath,\highdegreemath{\tau},\maxmatchmath, \conncompmath$, \ts{event-level privacy for output-determined algorithms is essentially as hard as item-level privacy}, thus, our algorithms are essentially optimal within the class of output-determined algorithms. Hence, if there exist algorithms that have a polynomially better accuracy for event-level, they must use more information about the input sequences, than just the true function values.

\begin{definition}[Output-determined algorithms~\cite{HenzingerSS24}]\label{def:output-determined}
Let $\gamma\in[0,1)$. Let $f$ be a function on $\uni^T$. An algorithm $\Alg:\uni^T\rightarrow \range(\Alg)$ is $(f,\gamma)$-\emph{output-determined} if, for every two sequences $\graphseq_1,\graphseq_2\in\uni^T$
satisfying $f(\graphseq_1)=f(\graphseq_2)$ and for all $\out\in\range(\Alg)$,  
\[\Pr[\mathcal{A}(\graphseq_1)\in \out]\leq \Pr[\mathcal{A}(\graphseq_2)\in \out]+\gamma.\]
\end{definition}
The lower bounds are proved via reductions from the 1-way marginals problem (see \Cref{def:marginals}).

\begin{theorem}\label{thm:general_lower_output} Let $f:\allgraphs{V}\rightarrow\R$ be an additive function with a $2$-edge distinguishing gadget of size $(n_g,m_g)$ and weight $w$ for some $n_g,m_g,w\in\N$. Then, for all $\eps\in(0,1]$ and $\gamma,\delta\in(0,1)$ with $\gamma\leq \delta$, and sufficiently large $T,N\in\N$, every $(f,\gamma)$-output-determined event-level \ededgedp algorithm which is $(\alpha,\frac{1}{3})$-accurate for $f$ on all dynamic graph sequences with $N$ nodes and length $T$ satisfies 
     \begin{itemize}
         \item if $\delta>0$ and $\delta=o\left(\frac{1}{N}\right)$, then $\alpha=\Omega\left(w\cdot\min\left(\frac{T^{1/3}}{\eps^{2/3}\log^{2/3} T},\frac{T}{m_g},\frac{N}{n_g}\right)\right)$.
         \item if $\delta=0$, then $\alpha=\Omega\left(\min\left(\sqrt{\frac{T}{\eps}},\frac{T}{m_g},\frac{N}{n_g}\right)\right)$.
     \end{itemize}
     \end{theorem}

\begin{proof} 
Let $n,d$ be in $\N$ to be fixed later. We reduce from $\marginals{n}{d}$. Let $H=(V',E')$ and $(H,e_1,e_2)$ be a 2-edge distinguishing gadget of size $(n_g,m_g)$ of weight $w$ for $f$. W.l.o.g.\ assume $e_1,e_2\notin E'$. Given an input $Y$ to $\marginals{n}{d}$, we define a graph sequence $\graphseq(Y)=(G_1,\dots G_T)$ with $n\cdot n_g$ nodes and length $T=(m_g+2d)n$ as follows: In the \emph{initialization phase}, we build $n$ copies of $H$, denoted $H^{(1)},\dots, H^{(n)}$, using $m_g n$ time steps. For a vertex pair $e$ in $H$ and $i\in[n]$, let $e^{(i)}$ denote the copy of $e$ in $H^{(i)}$. For all $i\in[n]$, we insert edge $e_1^{(i)}$ at time step $m_gn+i$. We set $t_0=(m_g+1)n$ and record $\Alg(\graphseq(Y))_{t_0}=a_0$.

For each $j\in[d]$, we first have an \emph{insertion phase},  where for all $i\in[n]$, if $Y_i[j]=1$, then we insert $e_2^{(i)}$ into $H^{(i)}$ at time $(m_g+1)n+2(j-1)n+i$; otherwise, we do nothing at that time step. We then set $t_j=(m_g+1)n+2(j-1)n+n$ and $\Alg(\graphseq(Y))_{t_j}=a_j$. We return $(a_j-a_0)/(wn)$ as an estimate for the $j$th marginal. For all but the last phase, the insertion phase is followed by a \emph{deletion phase}, where the edges that were inserted in the insertion phase are deleted.
    
     \paragraph{Privacy.} Let $Y'$ be an input to $\marginals{n}{d}$ with $Y\sim Y'$, such that $Y_{i^*}\neq Y'_{i^*}$ for some ${i^*}\in[n]$. Let $X$ be the dynamic graph sequence which is equal to $G(Y)$, except that the edge $e_1^{(i^*)}$ is never inserted. Then $\graphseq(Y)$ and $X$ are event-level  edge-neighboring. Similarly, let $X'$ be the same graph sequence as $\graphseq(Y')$, except that $e_1^{(i^*)}$ is never inserted. Then $G(Y')$ and $X'$ are event-level  edge-neighboring. Sequences $X$ and $X'$ differ only in edge insertions of $e_2^{(i^*)}$. However, for both $X$ and $X'$, the edge $e_1^{(i^*)}$ is always missing. Thus, $X$ and $X'$ produce the same output sequences for $f$. Then, for all $U\in\range(\Alg)$:
    \begin{align*}
        \Pr(\Alg(\graphseq(Y))\in U]&\leq e^{\eps}\Pr(\Alg(X)\in U)+\delta\\ &\leq e^{\eps}\Pr(\Alg(X')\in U)+\delta+\gamma\\ &\leq e^{2\eps}\Pr(\Alg(\graphseq(Y'))\in U)+\delta(1+e^{\eps})+\gamma.
    \end{align*}
    \sr{Since $\gamma\leq \delta$, the resulting algorithm for Marginals is $(2\eps,\delta(2+e^\eps))$-DP.}
    
     \paragraph{Accuracy.} Let $\graphseq(Y)=(G_1,\dots, G_T)$. At time $t_j$ in $\graphseq(Y)$,  both $e_1^{(i)}$ and $e_2^{(i)}$ are in $H^{(i)}$ if and only if $Y_i[j]=1$. Thus, $f(G_{t_j})=f(G_{t_0})+w\cdot \sum_{i\in[n]}Y_i[j]$. If $\Alg$ has error at most $\alpha$ at all time steps with probability at least $1-\beta$, then the resulting algorithm for marginals has error at most $\eta=\frac{2\alpha}{nw}$ with the same probability.
     
      \emph{Parameter settings.} 
{\bf Case $\delta>0$:} Given $T$ and $N$, set $d=\big\lfloor\left(T\eps\log{T\eps}\right)^{2/3}\big\rfloor$ and $n=\lfloor\min\left(\frac{\sqrt{d}}{6\eps\log d},\frac{T}{2m_g}, \frac{N}{n_g}\right)\rfloor$. The construction above 
   gives a sequence with at most $N$ nodes and $2nd+T/2$ updates, where 
   $$2nd\leq\frac{d^{3/2}}{3\eps\log d}\leq\frac{{T\eps}\log ({T\eps})}{3\eps\cdot\frac{2}{3}\log({T\eps}\log ({T\eps}))}\leq \frac T2.$$
   We can add  dummy nodes and dummy time steps to get a dynamic $N$-node graph sequence of length $T$. If we can estimate $f(G_t)$ with error up to $\alpha$ at all time steps with probability at least $2/3$, then for all $j\in[d]$, we can estimate the $j$th marginal given by $\frac{1}{n}\sum_{i\in n}Y_i[j]=\frac{f(G_{t_j})-f(G_{t_0})}{nw}$ up to  error $\frac{2\alpha}{nw}$ with the same probability. \Cref{lem:marginals}
    gives $\alpha=\Omega\left(wn\cdot\min\left(\frac{\sqrt{d}}{n\eps\log d},1\right)\right)=\Omega\left(w\cdot\min\left(\frac{T^{1/3}}{\eps^{2/3}\log^{2/3} T},\frac{T}{m_g}, \frac{N}{n_g}\right)\right)$.

    {\bf Case $\delta=0$:} Given  $T$ and $N$, set $d=\lfloor\sqrt{T\eps}\rfloor$ and $n=\lfloor\min\left(\frac{T}{(m_g+1)},\frac{N}{n_g}, \frac{d}{4\eps}\right)\rfloor$. The construction above gives a sequence with at most $N$ nodes and at most $T$ updates, and we can add dummy nodes and dummy time steps to get a dynamic $N$-node graph sequence of length $T$. By \Cref{lem:marginals}, we have 
    $\alpha=\Omega\left(wn\cdot\min\left(\frac{d}{n\eps},1\right)\right)=\Omega\left(w\cdot\min\left(\sqrt{\frac{T}{\eps}},\frac{T}{m_g},\frac{N}{n_g}\right)\right)$.
\end{proof}

\subsubsection{Output-Determined Lower Bounds for Specific Problems}\label{sec:lb-output-determined-specific-problems}

Using 
the 2-edge distinguishing gadgets from \Cref{fig:gadgets_gen} in \Cref{thm:general_lower_output}, we immediately get the following corollary for  
$\maxmatch$,$\conncomp$, and $\highdegree{1}$  from \Cref{thm:general_item_lower}. 
By the same reasoning as in the proof of Corollary~\ref{cor:lb-event-level}, this lower bound holds for \degreehist and $\highdegree{\tau}$ for $\tau\geq 2$.

\begin{corollary}[Output-determined event-level lower bounds]\label{cor:lb-event-level-outputd}
   Let $\tau\in\N$ and $f\in\{\maxmatchmath,\conncompmath,\highdegreemath{\tau},\degreehistmath\}$. Then, for all $\eps\in(0,1]$, $\gamma,\delta\in[0,1)$ with $\gamma\leq \delta$, and sufficiently large $T, N,D\in\N$, every $(f,\gamma)$-output-determined event-level, $(\eps,\delta)$-edge DP algorithm which is $(\alpha,\frac{1}{3})$-accurate  for $f$ on all dynamic graph sequences with $N$ nodes and length $T$ satisfies 
   \begin{enumerate}
       \item if $\delta>0$ and $\delta=o(\frac 1 T)$, then 
       $\alpha=\Omega\left(\min\left(\frac{T^{1/3}}{\eps^{2/3}\log^{2/3} T},T, N\right)\right)$.
       \item if $\delta=0$, then 
       $\alpha=\Omega\left(\min\left(\sqrt{\frac{T}{\eps}},T,N\right)\right)$.
   \end{enumerate}
\end{corollary}

\bibliographystyle{alpha}
\bibliography{bibliography}

\appendix

\section{Additional Privacy Background}
\paragraph{Continual release model.} Let $T\in \mathbb{N}$. In the continual release model, the input is given by a stream $x_1,\dots, x_T$, where $x_t\in \uni$ for all $t\in[T]$. 
An algorithm $\Alg$ in the continual release model receives at every time step $t\in[T]$ an input $x_t$ and produces an output $a^t=\Alg(x_1,\dots,x_t)$. 

\begin{definition}[Event-neighboring streams] Let $T\in\mathbb{N}$. Let $x=(x_1, \dots, x_T)$ with $x_t\in \uni$ for all $t\in[T]$ and $y=(y_1, \dots, y_T)$ with $y_t\in \uni$ for all $t\in[T]$ be two input streams. The streams $x$ and $y$ are \emph{event-neighboring}, denoted $x\sim y$, if there exists a time ${t^{*}}$ such that $x_t=y_t$ for all $t\neq{t^{*}}$.
\end{definition}

\begin{definition}[Event-level differential privacy] Let $T\in \mathbb{N}$ and $\Alg$ be an algorithm in the continual release model on $\mathbb{\uni}^{T}$. Let $\eps>0$ and $\delta\in [0,1)$. The algorithm $\Alg$ is \emph{event-level $(\eps,\delta)$-differentially private ($(\eps,\delta)$-DP)} if for all $\out\subseteq \mathrm{range}(\Alg^T)$ and all event-neighboring streams $x,y\in \uni^{T}$ we have
    \begin{align*}
        \Pr[(\Alg(x_1,\dots, x_t))_{t\leq T}\in  \out]\leq e^{\eps} \Pr[(\Alg(y_1,\dots, y_t))_{t\leq T}\in  \out]+\delta.
    \end{align*}
    If $\delta=0$ then $\Alg$ is \emph{event-level $\eps$-differentially private ($\eps$-DP)}.
    \end{definition}
    
    \begin{definition}[$(\alpha,\beta)$-accuracy] Let $k\in\N$ and $\uni$ be a universe of data items. Let $f$ be a function $\uni^*\rightarrow \mathbb{R}^k$. Let $x=(x_1,\dots,x_T)$ be a stream of elements from $\uni$. A continual release algorithm $\Alg$ is $(\alpha,\beta)$-accurate for estimating $f$ on $x$ if $\Pr\left[\max_{t\in[T]}\|\Alg(x_1,\dots,x_t)-f(x_1,\dots,x_t)\|_{\infty}>\alpha\right]\leq \beta.$
    \end{definition}

\subsection{Continual Histogram and Continual Counting}
\begin{definition}[Continual counting]
Let 
$T\in\N$. For an input stream $x=(x_1, \dots, x_T)$ with $x_t\in\{0,1\}$ for all $t\in[T]$, the \emph{continual counting problem} is to compute $\sum_{t'\in[t]}x_{t'}$ at every time step $t\in[T]$.
\end{definition}

\begin{definition}[$b$-bounded continual histogram]\label{def:bounded_hist}
Let $T,d,b\in \N$ with $b\leq d$. Let $\uni\subseteq \{-1,0,1\}^d$ be the set of vectors in $\{-1,0,1\}^d$ which have at most $b$ nonzero entries. For an input stream $x=(x_1, \dots, x_T)$ with $x_t\in\uni$ for all $t\in[T]$, the \emph{$b$-bounded continual histogram problem} is to compute $s_i^t=\sum_{t'\leq t}x_{t'}[i]$ for all $i\in[d]$ at every time step $t\in[T]$.
\end{definition}

The continual counting problem and continual histogram problem are well studied \cite{DworkNPR10,ChanSS11,DworkNRR15,FichtenbergerHU23,HenzingerUU23,CohenLNSS}. In the following lemma, the first bound follows from composing $d$ binary counters from \cite{DworkNPR10} and the analysis in \cite{ChanSS11}; the second bound (for $\delta>0$) is from \cite{FichtenbergerHU23}. Note that while solutions for continual counting are usually formulated for the universe $\{0,1\}$, inputs over $\{-1,0,1\}$ can be reduced to that case by running a separate counter for entries from $\{-1,0\}$ and $\{0,1\}$ and subtracting the first from the second.

\begin{lemma}[Algorithm for $b$-bounded continual histogram]\label{lem:dphist} Let $T,d,b\in\N$ with $b\leq d$, and let $\uni$ be as in Definition~\ref{def:bounded_hist}. Let $\eps\in(0,1)$ and $\delta\in[0,1)$. There exists an $(\eps,\delta)$-differentially private algorithm for $b$-bounded continual histogram which is $(\alpha,\beta)$-accurate for all input streams from $\uni^T$, where
\begin{enumerate}
    \item if $\delta=0$, then $\alpha=O\left(\frac{b}{\eps}\log^2\left(\frac{dT}{\beta}\right)\right)$.\label{lem:dphist_eps}
    \item if $\delta>0$, then $\alpha=O\left(\frac{\log T}{\eps}\left(\ln\frac{1}{\delta}\right)\sqrt{b\ln(dT)}\right).$\label{lem:dphist_epsdel}
\end{enumerate}
\end{lemma}
\end{document}